\documentclass[preprint,iop,amsmath,amssymb]{revtex4-1}
\usepackage{bm}
\usepackage{epsfig}
\usepackage{amsmath}
\usepackage{color}
\newcommand{\nix}[1]{}
\usepackage[unicode=true,colorlinks=true,urlcolor=blue,citecolor=blue]{hyperref}

\begin{document}

\title{Theory of an electron asymmetric scattering on skyrmion textures in two-dimensional systems}

\author{K.S. Denisov}
\email{denisokonstantin@gmail.com} 
\affiliation{Ioffe Institute, 194021 St.Petersburg, Russia}

\begin{abstract}
We discuss in details the electron scattering pattern on skyrmion-like magnetic textures in two-dimensional geometry. 
The special attention is focused on analyzing the 
scattering asymmetry, which is a precursor of the topological Hall effect. 
We present analytical results valid for the limiting regimes of strong and weak coupling, we also describe the numerical scheme that gives access to the exact solution of the scattering problem. Based on the numerical computations we 
investigate the properties of the asymmetric scattering for an arbitrary magnitude of the interaction strength and 
the topology of a magnetic texture. 
We analyze in details the conditions when the topological charge of a magnetic texture is indeed essential for the appearance of the Hall response.
We~also describe how the electron scattering asymmetry is modified due to an additional short-range impurity located inside a magnetic skyrmion.

\end{abstract}

\date{\today}

\maketitle

\section{Introduction}



\vspace{-0.2cm}

Rapidly growing physics of the topological magnetic textures~\cite{NagaosaNature,yu2018transformation,fert2017magnetic,wiesendanger2016nanoscale} 
pays a special attention to the emerging nontrivial electrodynamics~\cite{sundaram1999wave,Berry1,buhl2017topological} and to the topological Hall effect~(THE) in particular. 
Introduced originally to describe 
the Hall effect in a skyrmion lattice~\cite{BrunoDugaev} 
and being experimentally evidenced in such systems~\cite{MnSiAPhase,leroux2018skyrmion,kanazawa2011large,B20_FeCoGe}, 
the THE contribution however is not specific to this unique kind of geometry. 
The recent advances in imaging techniques have revealed different material platforms possessing individual skyrmions with size ranging from sub-$100$ nm~\cite{moreau2016additive,legrand2017room,raju2019evolution,soumyanarayanan2017tunable,wang2018ferroelectrically}
down to sub-$10$ nm~\cite{Romming2013,romming2015field,meyer2019isolated}. 
While 
the subsequent transport studies of both the discretized skyrmion geometry~\cite{zeissler2018discrete,DiscretHall,maccariello2018electrical} and the non-regular skyrmion arrays~\cite{raju2019evolution,EuO,EuO-2,THE_TI,karube2018disordered,wang2018ferroelectrically,meng2019observation} 
in general confirm the presence of THE signatures, 
the estimation of THE magnitude in these systems is still debated as it can be dramatically modified due to 
an additional impurity scattering or the nonadiabaticity of a carrier spin motion~\cite{nakazawa2018topological,nakazawa2018weak,denisov2018general,ishizuka2018impurity,ishizuka2018spin}.
Furthermore, accounting for a mixed spin-orbital electron dynamics becomes especially important when 
one reduces the skyrmion size and thus inevitably enters into the clean limit of a ballistic electron motion inside the skyrmion. 
One common approach capable for analyzing these issues is based on the tight-binding model, there are various simulations of THE for individual magnetic textures~\cite{Metalidis,TaillefumierPRB2006,Ohe-PRB2007,Arab_Papa,Akosa_Tret} and skyrmion latices~\cite{QTHE,Gobel,gobel2018}. 
However, as an alternative 
and somewhat more flexible description 
one has recently appealed to the scattering theory~\cite{denisov-SciRep,araki2017skyrmion}. 
Analyzing an electron scattering pattern 
has allowed to quantify 
the renormalization of THE magnitude 
due to various factors 
on a unified platform~\cite{denisov-prl2016,denisov-SciRep,denisov2018general} 
thus revealing 
a high usefulness of this approach. 

In this manuscript we present the comprehensive investigation of the electron scattering 
on skyrmion-like textures. 
Our main purpose is to analyze the properties of 
an asymmetric scattering for an arbitrary coupling strength including  weak and strong coupling regimes, 
as well as to clarify the conditions when the topology of a magnetic texture is essential. 
Encouraged by experimental indications that skyrmions tend to be captured by structural defects~\cite{meyer2019isolated,raju2019evolution} we also analyze how an additional impurity potential affects the Hall response. 

The paper is organized as follows. 
In Sec.~\ref{s_Frame} we introduce the framework for the electron scattering on a skyrmion texture. 
In Sec.~\ref{s2} we develop the analytic descriptions for the weak and strong coupling regimes; 
the numerical scheme is further developed in Sec.~\ref{s4} to address the exact solution of the scattering problem. 
Based on the numerical calculations we analyze various scattering scenarios in Sec.~\ref{s5}. 
In Sec.~\ref{sAdiab},~\ref{s-Quasi} and~\ref{sOne} we carefully examine the role of the skyrmion topology on the Hall current. 
The scattering on electrically charged skyrmions is discussed in Sec.~\ref{sWeak} and Sec.~\ref{sCharge}. 
The scattering features driven by the nonadiabaticity of the electron spin motion can be found in~\ref{sWeak},~\ref{sSkyrmion}. 



\section{Scattering framework}

\label{s_Frame}

\subsection{Skyrmion scattering potential}

\begin{figure}
	\centering	
	\includegraphics[width=0.45\textwidth]{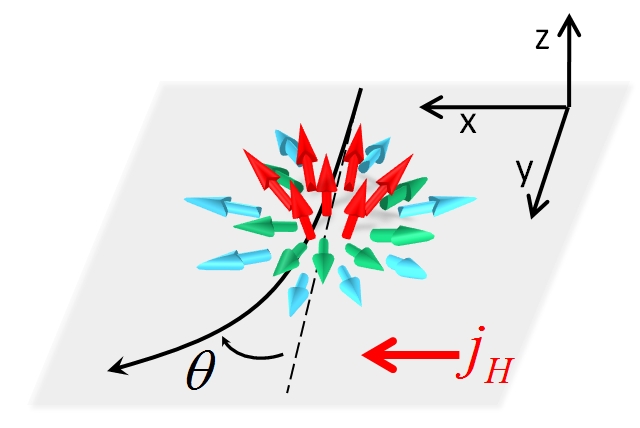}
	\qquad
	\includegraphics[width=0.45\textwidth]{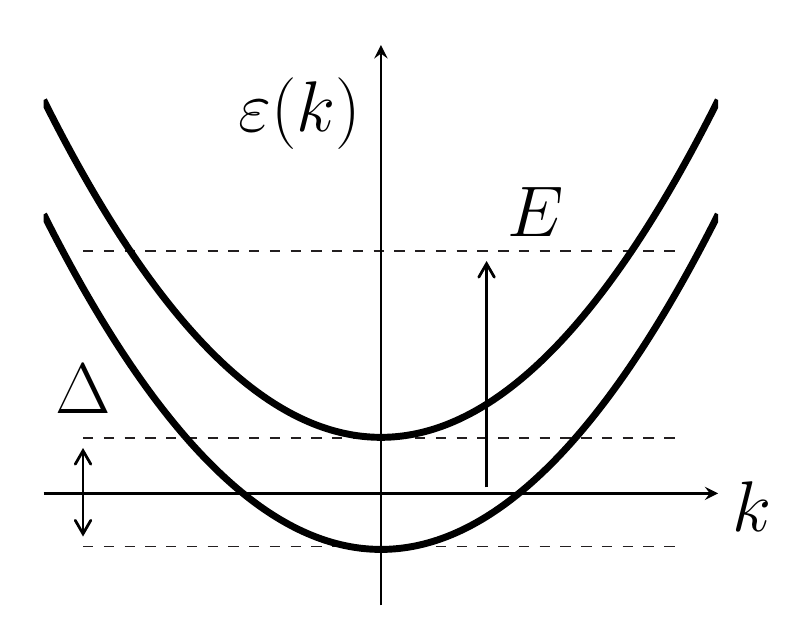}	
	\caption{Electron scattering on a magnetic texture and the electron free motion spectrum.}	
	\label{f2}
\end{figure}


We consider a 2D system with the electron Hamiltonian $\mathcal{H}$ 
given by:
\begin{align}
& \mathcal{H} = \mathcal{H}_0 + V(\boldsymbol{r}),
\hspace{1cm}
\mathcal{H}_0 = \frac{\hat{\bm{p}}^2}{2m_0} - \frac{\Delta}{2}\sigma_z, 
\label{eq_First}
\end{align}
here $\mathcal{H}_0$ describes the 
electron free motion, 
$m_{0}$ is an effective mass, $\hat{\bm{p}}$ is the momentum operator, $\Delta>0$ is the spin splitting of the electron subbands, 
the $\boldsymbol{r}$-dependent term $V(\boldsymbol{r})$ 
is a scattering potential. 
The spectrum shown in Fig.\ref{f2} consists of two parabolas shifted by $\Delta$, the energies are given by 
$
\varepsilon_{\bm{k}}^s = {\hbar^2 \bm{k}^2}/{2m_0} - s {\Delta},
$
where
$s=\pm {1}/{2}$ is the electron spin projection onto $z$ axis. 
There are two regimes with respect to the position of the electron energy $E$ (see Fig.~\ref{f2}), either 
$E>\Delta/2$ so both spin subbands are available for a free motion, or $E< \Delta/2$ and the propagation in the spin-down subband is suppressed. 
In what follows we denote the wavevectors at energy $E$  according to the following notation: 
\begin{align}
& (E>\Delta/2) \hspace{1cm} k_s = \sqrt{2m_0(E + s \Delta)}, 
\qquad
k = \sqrt{2m_0 E}. 
\label{eq_wvNom}
\end{align}

In this paper we consider the scattering potential $V(\boldsymbol{r})$ of the following form:
\begin{equation}
V(\boldsymbol{r}) = -g \begin{pmatrix}
v_1(r)
& u(r) e^{-i\left( \chi \phi + \gamma \right)}
\\
u(r) e^{i\left( \chi \phi + \gamma \right)}
&
v_2(r)
\end{pmatrix},
\label{eq_Vchiral}
\end{equation}
where $\boldsymbol{r}=(r,\phi)$, $g$ is a coupling constant, $v_{1,2}(r)$ and $u(r)$ 
are dimensionless real functions of $r$ only, 
the parameter $\chi = \pm 1, \pm 2, \cdots$ takes integer values,  
$\gamma$ is an arbitrary phase. 
We will assume that the potential has a localized character so that $v_{1,2},u \to 0$ at $r \gtrsim  r_0$, where $r_0$ is a localization radius.  
Of key importance is the dependence of ${V}(\boldsymbol{r})$ on the polar angle $\phi$ entering in its off-diagonal components. The $\phi$-dependence leads to the non-commutativity of the Hamiltonian with the operator of angular momentum $-i \partial_{\phi}$, 
as a result 
the electron scattering on $V(\bm{r})$ gets an asymmetric character. 
A comprehensive description of this phenomenon is the main subject of the present paper. 

Let us comment on physics underlying the chosen form of $V(\boldsymbol{r})$. 
This type of potentials is relevant for magnetic materials when an electron interacts with a single chiral spin texture, such as magnetic skyrmion. 
Let $\bm{n}(\boldsymbol{r})$ be a unit vector directed along the local magnetization. 
For an individual chiral spin texture $\bm{n}(\bm{r})$  can be generally written~as: 
\begin{equation}
\bm{n}(\boldsymbol{r}) = 
\Bigl(
n_{\parallel}(r) \cos{\left( \chi \phi + \gamma \right)},
n_{\parallel}(r) \sin{\left( \chi \phi + \gamma \right)},
n_{z}(r) 
\Bigr),
\end{equation}
where ($\chi, \gamma$) correspond to the vorticity and the helicity of the spin texture, respectively, 
and $n_{z,\parallel}(r)$ describe the radial profiles 
(it is assumed that at $r \gtrsim r_0$ one has $n_z \to 1, n_\parallel \to 0$). 
The scattering potential $V(\bm{r})$ from Eq.~\ref{eq_Vchiral} appears due to an electron exchange interaction with a static magnetization field of this shape. 
If no other perturbation is present we can relate $g$ to an exchange interaction constant and the functions $v_{1,2}(r),u(r)$ to the spin profiles:
\begin{equation}
v_{1}(r)=-v_2(r) = n_z(r)-1,
\qquad
u(r) = n_{\parallel}(r).
\label{eq_ab}
\end{equation}
The uniform background component 
outside the texture core $\bm{n}({r} \gtrsim r_0) = \bm{e}_z$ gives rise to the spin subband splitting $\Delta = 2g$ in this case. 

We shall mention that when an electron
scattering is induced entirely by the perturbation of the magnetization 
one should assume an additional 
coupling $v_1 = -v_2$ between the diagonal components of $V(\bm{r})$. 
In what follows, however, we will develop the theory 
with no restrictions on $v_{1,2}(r)$ functions. 
By making this generalization 
we can also introduce to our consideration the possibility of 
a scalar potential $U_0$ to be superimposed on a spin texture. This situation has a great practical interest, as there are numerous experimental observations that skyrmions tend to be pinned by structural defects, i.e. by charged impurities. 
In particular, 
the approximation $v_1 \approx v_2$ instead of $v_1 = - v_2$ can be considered to describe the extreme regime with $U_0 \gg g$. 
The theory present in this work allows us to analyze both these cases.

\subsection{Scattering rates}

In this work we treat the scattering problem using the $\hat{T}(z)$-operator which satisfies the Lippman-Schwinger equation:
\begin{equation}
\label{eq_LPSH}
\hat{T}(z) = {V} + V \hat{G}_0(z) \hat{T}(z),
\end{equation}
where $\hat{G}_0(z) = (z- \mathcal{H}_0)^{-1}$ is the Green operator corresponding to the free Hamiltonian, and $V$ corresponds to the scattering potential $V(\bm{r})$ defined in Eq.~\ref{eq_Vchiral}. 
Since $V(\bm{r})$ is a $2\times 2$ matrix 
there are generally 4 scattering channels. 
To describe an elastic electron scattering with the energy $E$  from $(\bm{k}',s')$ to $(\bm{k},s)$ states one deals with the $T$-matrix on a mass shell:
\begin{equation*}
T_{\bm{kk}'}^{ss'} \equiv \lim_{\delta \to 0} \langle 
\bm{k}s | \hat{T}(E+i\delta) | \bm{k}'s'
\rangle. 
\end{equation*}
The square modulus of so defined $T$-matrix elements $|T_{\bm{kk}'}^{ss'}|^2$ determine the scattering rates. 
In particular, the differential scattering cross-section in 2D geometry is defined as~\cite{adhikari1986quantum}:
\begin{equation}
\label{eq_Cross_For_AD}
\frac{d\sigma_{ss'}}{d\theta} = \frac{m_0^2}{2\pi \hbar^4 k_{s'}} |T_{\bm{kk}'}^{ss'}|^2,
\end{equation}
where $\theta$ is the scattering angle, i.e. the angle between $\bm{k}$ and $\bm{k}'$. 
In what follows, however, we will describe the scattering 
using the  
symmetric 
$\mathcal{G}_{{\bm{kk}}'}^{ss'} = \mathcal{G}_{\bm{k'k}}^{ss'}$, 
and asymmetric 
$\mathcal{J}_{\bm{kk}'}^{ss'} = -\mathcal{J}_{\bm{k'k}}^{s's}$ 
dimensionless functions defined as:
\begin{equation}
\label{eq_T-nu}
|T_{\boldsymbol{kk}'}^{ss'}|^2 = \frac{1}{\nu_0^2} \left( \mathcal{G}_{\bm{kk}'}^{ss'} + \mathcal{J}_{{\bm{kk}}'}^{ss'} \right), 
\qquad
\nu_0 = \frac{m_0}{2\pi \hbar^2}. 
\end{equation}
Here the prefactor corresponds to the two-dimensional density of states $\nu_0$.  
The reason to extract 
$\nu_0^2$ explicitly from $|T_{\bm{kk}'}^{ss'}|^2$ becomes clear when using the classical description present in~\ref{sAdiab}; 
it also allows for a compact and natural representation of the Hall resistivity, see the details in Ref.~\cite{denisov2018general}.

The asymmetric terms $\mathcal{J}_{\bm{kk}'}^{ss'}$ leading to the Hall response appear due to the intrinsic angular asymmetry of the considered chiral potentials. 
The integral quantities describing the transverse currents are given by:
\begin{equation}
\mathcal{J}_{ss'} = \int\limits_0^{2\pi} \mathcal{J}_{\bm{kk}'}^{ss'} \sin{\theta} d\theta. 
\label{eq_J_TOTAL}
\end{equation}

In the following sections we calculate $\mathcal{G}_{\bm{kk}'}^{ss'}, \mathcal{J}_{\bm{kk}'}^{ss'}$ and consider the properties of an electron scattering 
in various regimes and for different potential shapes. 

\section{Analytical results}

\label{s2}

Let us comment on some general features associated with the scattering on chiral potentials $V(\bm{r})$. 
There are two types of dynamic processes taking place during the electron motion inside the scattering region. 
Firstly, there is an evolution of its orbital trajectory associated with the change of the momentum in the given potential. 
Secondly, there is an electron spin rotation driven by its coupling with the spatially non-homogeneous magnetization field. 
Of highest importance is that these two processes affect each other 
leading to the appearance of the scattering asymmetry. 

In order to develop an \textit{analytical} description for some limiting regimes we should distinguish the role of parameters affecting both orbital and spin motions. 
For instance, assuming $g={const}$ 
the magnitude of product $k r_0$ would determine two orbitally different regimes. Namely, there is 
the transition from quantum isotropic scattering (described perturbatively) at $k r_0 \lesssim 1$ to the quasiclassical low-angle motion at $k r_0 \gg 1$ governed by the Newton mechanics. 
The character of an electron spin motion in its turn changes essentially depending on the magnitude of the adiabatic parameter determined as $\lambda_a = \omega_{ex} \tau_{\rm fly}$, here $\omega_{ex}=2g/\hbar$ corresponds to the energy difference between spin up and spin down states, and 
$\tau_{\rm fly} = (2r_0)/ v$ is the time of electron presence inside a texture core, here $v = \sqrt{2E/m_0}$ is the electron velocity. 
The magnitude of $\lambda_a$ shows if the electron has enough time for its spin to become adiabatically co-aligned with the local magnetization direction ($\lambda_a \gg 1$),  or the perturbation is rather instantaneous ($\lambda_a \lesssim 1$) so that after flying out of the potential region the electron spin only experiences a small rotation with respect to its initial direction. 
Naturally, the latter scenario $\lambda_a \lesssim 1$ is accompanied by the activation of the spin-flip scattering channels; the adiabatic regime on the contrary is featured by the suppression of the spin-flip processes. 



The adiabatic parameter can be written in form $\lambda_a = (2g/E) \cdot (k r_0)$, which indicates that 
the change of the potential radius $r_0$ would affect both the orbital and the spin motions at the same moment.  
As a result it is reasonable to treat 
the so-called weak coupling regime ($\lambda_a \lesssim 1$) along with the condition $k r_0 \lesssim 1$ 
on the basis of the perturbation theory, 
while to consider the opposite adiabatic regime 
assuming $k r_0 \gg 1$ and 
using the quasiclassical approximation. 
Below in this section we present the analytical results for these two limiting regimes.
The numerical scheme is further developed in~\ref{s4} to address the exact solution of the scattering problem. 


\subsection{Perturbation theory}

\label{sWeak}

In this section we will get analytical expressions for the scattering rates $\mathcal{G}_{\bm{kk}'}^{ss'}, \mathcal{J}_{\bm{kk}'}^{ss'}$ 
using the perturbation theory. 
We assume $\lambda_a \lesssim 1$ and that 
the scattering potential has a short-range character, which means that $V(\bm{r})$ is 
nonzero only at 
$k r_0 \lesssim 1$.
The starting point is 
Eq.~\ref{eq_LPSH} written for the $T$-matrix on a mass shell:
\begin{equation}
T_{\boldsymbol{kk}'}^{ss'} = {V}_{\boldsymbol{kk}'}^{ss'} + \sum_{\bm{g},s''} \frac{{V}_{\boldsymbol{kg}}^{ss''} {T}_{\boldsymbol{gk}'}^{s''s'}}{E - \varepsilon_{\boldsymbol{g}}^{s''} + i0},
\end{equation}
where $E$ is the electron energy, $\varepsilon_{\boldsymbol{k}}^s$ is the band spectrum given by Eq.~\ref{eq_First}, and ${V}_{\boldsymbol{kk}'}^{ss'}$ is the matrix element of the scattering potential:
\begin{align}
& {V}_{\boldsymbol{kk}'}^{ss'} = - g
\begin{pmatrix}
v_1(q) & -i e^{-i(\chi \varphi_q + \gamma)} u(q)
\\
-i e^{i(\chi \varphi_q + \gamma)} u(q) & v_2(q)
\end{pmatrix},
\label{eq_Vm}
\\
&v_{1,2}(q) = 2\pi \int\limits_0^{\infty} rdr J_0(qr) v_{1,2}(r), 
\hspace{0.4cm}
u(q) = 2\pi \int\limits_0^{\infty} rdr J_{1}(qr) u(r), 
\notag
\end{align}
where $\boldsymbol{q} = \boldsymbol{k}- \boldsymbol{k}' \equiv (q,\varphi_q)$, $\varphi_q$ is the polar angle of $\boldsymbol{q}$,  $J_0,J_1$ are the Bessel functions of zero and first kind respectively (in this section we consider only $\chi = \pm 1$). 
We are focused on a short-range scattering potential $k r_0 \lesssim 1$, at that one can replace the Bessel functions in the matrix elements Eq.~\ref{eq_Vm} by the approximations at small arguments $J_0(x) \approx 1, J_1(x) \approx x/2$:
\begin{align}
& v_{1,2}(q) = 2\pi r_0^2 \cdot \mathcal{I}_{\uparrow, \downarrow},
\qquad
u(q) = \pi r_0^3 q \cdot \mathcal{I}_{\parallel},
\notag
\\
& \mathcal{I}_{\uparrow, \downarrow} = \int\limits_0^1 v_{1,2}(x r_0) xdx, 
\qquad 
\mathcal{I}_{\parallel} = \int\limits_0^1 u(x r_0) x^2dx, 
\label{eq_ab}
\end{align}
where the dimensionless numbers $\mathcal{I}_{1,2,\parallel}$ are determined only by particular profiles $v_{1,2}, u$ and have no dependence on $r_0$ (at least in the limit $kr_0 \lesssim 1$). 
The $q$-dependent prefactor in $u(q)$ can be eliminated using the following expression for the exponent~$e^{\pm i \varphi_q}$:
\begin{equation}
\label{eq_Phi-q}
e^{\pm i \varphi_q} = \frac{1}{q} \left( ke^{\pm i\varphi} - k' e^{\pm i \varphi'} \right),
\end{equation}
where $\varphi,\varphi'$ are the polar angles of $\boldsymbol{k,k}'$ respectively.

We firstly consider the regime when both energy branches are available ($\Delta< 2E$). 
We will also approximate $k_{\uparrow} \approx k_{\downarrow} \equiv k$
assuming small spin splitting $\Delta/2E \ll 1$. 
In the lowest order of the perturbation theory the $T$-matrix is simply given by the matrix element $ T_{\boldsymbol{kk}'}^{ss'} = {V}_{\boldsymbol{kk}'}^{ss'}$. 
The asymmetric rates $\mathcal{J}_{\bm{kk}'}^{ss'}$ are absent in this approximation; 
the symmetric parts 
are given~by: 
\begin{align}
& 
 \mathcal{G}_{\bm{kk}'}^{\uparrow \uparrow} 
= \left(\frac{g}{2E}\right)^2 \left(k r_0\right)^4
\mathcal{I}_{\uparrow}^2 ,
\qquad
  \mathcal{G}_{\bm{kk}'}^{\downarrow \downarrow}
= \left(\frac{g}{2E}\right)^2 \left(k r_0\right)^4
\mathcal{I}_{\downarrow}^2
\notag
\\
& \mathcal{G}_{\bm{kk}'}^{\uparrow \downarrow} = \mathcal{G}_{\bm{kk}'}^{\downarrow \uparrow}=  \left(\frac{g}{2E}\right)^2 \left(k r_0\right)^6 \mathcal{I}_\parallel^2 \cdot
\sin^{2}{\theta}. 
\label{eq_GsymWL}
\end{align}
The scattering in the spin-conserving channels is isotropic 
in analogy with the $s$-scattering on $\delta$-potential. 
The scattering in the spin-flip channels, however, acquires an angular dependence featured by the suppression of the forward scattering. 
Let us mention that the spin-flip scattering rates contain an additional smallness due to the higher order of $k r_0$. 


We proceed with calculating the asymmetric terms $\mathcal{J}_{\bm{kk}'}^{ss'}$. 
The second-order correction to the $T$-matrix 
is given by:
\begin{align}
&\delta T_{\boldsymbol{kk}'}^{ss'} =  \sum_{\boldsymbol{g},s''} \frac{{V}_{\boldsymbol{kg}}^{ss''} {{V}}_{\boldsymbol{g k}'}^{s''s'}}{E - \varepsilon_{\boldsymbol{g}}^{s''} + i0} = 
 \mathcal{P} \sum_{\boldsymbol{g},s''} 
  \frac{{V}_{\boldsymbol{kg}}^{ss''} {{V}}_{\boldsymbol{gk}'}^{s''s'}}{E - \varepsilon_{\boldsymbol{g}}^{s''}} - i \pi \sum_{\boldsymbol{g},s''} \delta\left( E - \varepsilon_{\boldsymbol{g}}^{s''} \right) {V}_{\boldsymbol{kg}}^{ss''} {{V}}_{\boldsymbol{gk}'}^{s''s'}. 
  \label{eq_T2}
\end{align}
The first term in Eq.~\ref{eq_T2} is  
the correction to $\mathcal{G}_{\bm{kk}'}^{ss'}$ and we will neglect it (here $\mathcal{P}$ stands for the principal value). The second term 
gives rise to $\mathcal{J}_{\bm{kk}'}^{ss'}$ 
via the interference with the first Born approximation terms ${V}_{\boldsymbol{kk}'}^{ss'}$ 
in the square modulus of $T$-matrix:
\begin{align}
& \mathcal{J}_{\bm{kk}'}^{ss'} = \nu_0^3 \sum_{s''} \int\limits_0^{2\pi} d\varphi_g \cdot {\rm Im}\left[ {{V}}_{\boldsymbol{kg}}^{ss''} {{V}}_{\boldsymbol{gk}'}^{s''s'} {{V}}_{\boldsymbol{k}'\boldsymbol{k}}^{s's} \right],
\label{eq_J_pert}
\end{align}
here $\nu_0$ is defined in Eq.~\ref{eq_T-nu}. 
Indeed, upon the the replacement of the initial $(\boldsymbol{k}',s')$ and final $(\boldsymbol{k},s)$ scattering states the terms $\mathcal{J}_{\bm{kk}'}^{ss'}$ change their signs:
\begin{equation*}
\mathcal{J}_{\bm{kk}'}^{ss'} = \nu_0^3 \sum_{s''} \int\limits_0^{2\pi} d\varphi_g \cdot {\rm Im}\left[ {{V}}_{\boldsymbol{kg}}^{ss''} {{V}}_{\boldsymbol{gk}'}^{s''s'} {{V}}_{\boldsymbol{k}'\boldsymbol{k}}^{s's} \right]  = 
 \nu_0^3 \sum_{s''} \int\limits_0^{2\pi} d\varphi_g \cdot {\rm Im}\left[ \left({{V}}_{\boldsymbol{gk}}^{s''s} {{V}}_{\boldsymbol{k'g}}^{s's''} {{V}}_{\boldsymbol{k}\boldsymbol{k}'}^{ss'} \right)^\ast \right]  = - \mathcal{J}_{\bm{k'k}}^{s's}.
\label{eq_Asym1}
\end{equation*}
The diagonal matrix elements $V_{\bm{kk}'}^{\uparrow \uparrow}, V_{\bm{kk}'}^{\downarrow \downarrow}$ are real, so the imaginary part of the product~\ref{eq_J_pert} is nonzero only for 
the interference between one spin-conserving and two spin-flip scattering processes, i.e. 
${\rm Im}\left[ V_{\bm{k'k}}^{\uparrow \uparrow} V_{\bm{kg}}^{\uparrow \downarrow} V_{\bm{g k}'}^{\downarrow\uparrow} \right] \neq 0$ for $(\uparrow \uparrow)$ spin-conserving scattering channel, and 
${\rm Im}\left[V_{\bm{k'k}}^{\downarrow \uparrow} V_{\bm{kg}}^{\uparrow \downarrow} V_{\bm{g k}'}^{\downarrow \downarrow}
+
V_{\bm{k'k}}^{\downarrow \uparrow} V_{\bm{kg}}^{\uparrow \uparrow} V_{\bm{g k}'}^{\uparrow \downarrow} 
\right] \neq 0$ for $(\uparrow \downarrow)$ spin-flip scattering channel. 
Taking into account the explicit forms of $v_{1,2}(q), u(q)$ from Eq.~\ref{eq_ab} we get finally for $\mathcal{J}_{\bm{kk}'}^{ss'}$: 
\begin{align}
& \mathcal{J}_{\bm{kk}'}^{\uparrow \uparrow} =\zeta_{\uparrow}
\cdot \sin{\left(\chi \theta\right)}, 
\qquad
\mathcal{J}_{\bm{kk}'}^{\downarrow \downarrow} = - \zeta_{\downarrow}
\cdot \sin{\left(\chi \theta\right)}, 
\qquad
\mathcal{J}_{\bm{kk}'}^{\uparrow \downarrow}
= \mathcal{J}_{\bm{kk}'}^{\downarrow \uparrow}
=  \mathcal{J}_{\bm{kk}'}^{\uparrow \uparrow} + \mathcal{J}_{\bm{kk}'}^{\downarrow \downarrow} 
\notag
\\
& 
\zeta_{\uparrow, \downarrow} = \frac{\pi}{2} \left(\frac{g}{2E}\right)^3 {\left(k r_0\right)^8} \left(\mathcal{I}_\parallel^2 \cdot \mathcal{I}_{\uparrow, \downarrow} \right) 
\label{eq_J_cons1}
\end{align}

Let us discuss the main featurers of the obtained results. 
As one naturally expects for a short-range potential 
the angular dependence of $\mathcal{J}_{\bm{kk}'}^{ss'} 
$
is determined by the lowest asymmetric angular harmonic of the scattering angle $\theta$, i.e. by the $\sin{\theta}$. 
The magnitude of the asymmetric terms scales as the third order of the coupling constant $(g/E)^3$ and more remarkably by the eighth order of $(kr_0)^8$. 

It is especially important to analyze the spin-dependent properties of $\mathcal{J}_{\bm{kk}'}^{ss'}$.  
As follows from Eq.~\ref{eq_J_cons1} the character of the scattering asymmetry (charge or spin) essentially depends on 
whether the scattering potential $V(\bm{r})$ describes a purely 
magnetic texture ($v_1 = - v_2$), or it also contains an additional scalar potential ($v_1 \neq - v_2$). 
Indeed, for a pure chiral spin texture we have $\zeta_\uparrow = - \zeta_\downarrow$, which leads to the following coupling between $\mathcal{J}_{\bm{kk}'}^{ss'}$:
\begin{equation}
\mathcal{J}_{\bm{kk}'}^{\uparrow \uparrow} = \mathcal{J}_{\bm{kk}'}^{\downarrow \downarrow}, \qquad
\mathcal{J}_{\bm{kk}'}^{\uparrow \downarrow} = 2 \mathcal{J}_{\bm{kk}'}^{\uparrow \uparrow},
\hspace{2cm}
(v_1 = - v_2).
\label{eq_J_WK_Charge}
\end{equation}
These relations indicate that the spin-up and spin-down electrons are asymmetrically scattered in the same transverse direction; which one is determined by the product $(\chi \cdot n_z)$. 
This process leads to the appearance of the Hall current even for totally unpolarized carriers. 

On the contrary, if we consider that a strong scalar perturbation is superimposed on a chiral spin texture $U_0 \gg g$ 
and assume
that $v_1 \approx v_2$, the coupling between different scattering channels will take form:
\begin{equation}
\mathcal{J}_{\bm{kk}'}^{\uparrow \uparrow} \approx - \mathcal{J}_{\bm{kk}'}^{\downarrow \downarrow}, \qquad
\mathcal{J}_{\bm{kk}'}^{\uparrow \downarrow} \approx 0,
\hspace{2cm}
(v_1 \approx v_2).
\label{eq_J_WK_Spin}
\end{equation}
Thus a strong electron interaction with a nonmagnetic component of the scattering potential 
leads to the transverse spin current.  
Moreover, the direction of this current 
is sensitive to whether the nonmagnetic impurity is positively or negatively charged. 
In this regard 
a magnetic skyrmion electrostatic environment can dramatically affect the symmetry properties of the topological Hall effect and to significantly modify its dependence on a carrier spin polarization. 

The spin-dependent structure of the asymmetric scattering 
can be understood based on the spin chirality arguments. 
In case of a pure magnetic texture the Hall response is driven by the scalar spin chirality $\bm{M}_1 \cdot \left[ \bm{M}_2 \times \bm{M}_3 \right] $~\cite{Tatara,lazuta1978neutron,udalov2014hall}
which is irrelevant to 
the electron spin state; 
naturally it leads to the spin-independent skew scattering~\cite{denisov-prl2016}. 
On the contrary, when the scalar potential is present 
the scattering asymmetry 
can be induced 
due to the mixed product  
$\bm{S} \cdot \left[ \bm{M}_2 \times \bm{M}_3 \right] $~\cite{ishizuka2018impurity}
composing of both the magnetization vector spin chirality $\left[ \bm{M}_2 \times \bm{M}_3 \right] $ and the electron spin $\bm{S}$, this mechanism thus gives rise to the spin Hall effect.



Let us further consider the 
case when only one (spin-up) subband is activated ($\Delta > 2E$). 
The second-order correction to the $T$-matrix relevant for the scattering asymmetry is written:
\begin{equation}
\delta T_{\bm{kk}'}^{\uparrow \uparrow} = -i \nu_\uparrow^2(E) \int\limits_0^{2\pi} d\varphi_g \cdot \nu_{\downarrow}(E) V_{\bm{kg}}^{\uparrow \downarrow} V_{\bm{gk}'}^{\downarrow \uparrow}
\label{eq_Dissap}
\end{equation}
where $\nu_{\uparrow, \downarrow}(E)$ is the density of states in the corresponding spin subband. 
Since the energy $E$ lies below the bottom of $\varepsilon_{\downarrow}(\bm{k})$ spectrum we have $\nu_{\downarrow}(E) = 0$, which leads to the dissapearance of $\delta T_{\bm{kk}'}^{\uparrow \uparrow}$  and of the asymmetric scattering $\mathcal{J}_{\bm{kk}'}^{\uparrow \uparrow}=0$ correspondingly. 
Therefore the topological Hall effect is strongly suppressed 
in the regime $2E<\Delta, kr_0 \lesssim 1$. 


\subsection{Classical scattering and the adiabatic spin motion}

\label{sAdiab}

In this section we derive the expressions for the total asymmetric rates $\mathcal{J}_{ss'}$ from Eq.~\ref{eq_J_TOTAL} 
valid in the classical and adiabatic limits. 
In other words we assume that the scattering potential radius significantly exceeds the electron wavelength $k r_0 \gg 1$
and that the electron spin quantization axis is adiabatically co-aligned with local magnetization ($\lambda_a \gg 1$). 
In this case the scattering problem finds a
rather elegant solution based on classical mechanics.  

\begin{figure}[t]
	\centering
	\includegraphics[width=0.45\textwidth]{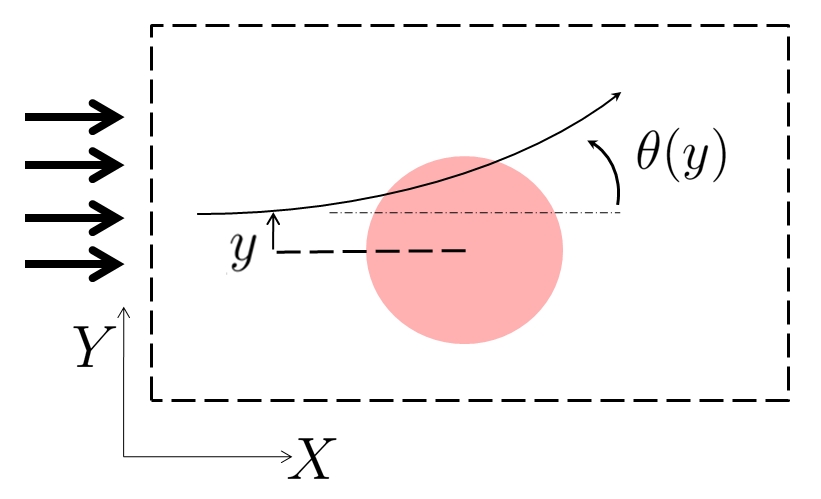}%
	\caption{The classical picture of an electron flux scattering.} 
	\label{f-scat}
\end{figure}

As a starting point 
we use Eq.~\ref{eq_Cross_For_AD} and express
$\mathcal{J}_{ss}$ 
for the spin conserving scattering channels
(the spin-flip scattering is suppressed $\mathcal{J}_{\uparrow \downarrow} \approx 0$
in the adiabatic limit) 
via the spin-dependent differential scattering cross-sections  $d\sigma_s/d\theta$:
\begin{equation}
\label{eq_J_Cross}
\mathcal{J}_{ss} = \frac{p_s}{2\pi \hbar} \int\limits_0^{2\pi} \frac{d\sigma_s}{d\theta} \sin{\theta} d\theta,
\end{equation}
here $\theta$ is the scattering angle, $p_s = \hbar k_s$ is the spin-dependent momenta from Eq.~\ref{eq_wvNom}.
In the 
assumed approximation 
we are allowed to calculate 
$d\sigma_s/d\theta$
by accounting for the classical trajectories of the electrons 
moving initially as a
uniform incident beam, see Fig.~\ref{f-scat}.
An electron 
approaching 
from the left boundary  
and having 
the impact parameter $y$ 
is deflected by the scattering angle $\theta(y)$ being the function of $y$, 
the electron momentum $\bm{p}_s$ after flying out of the scattering region 
has both projections 
$p_x^s = p_s \cos(\theta(y)), p_y^s = p_s \sin(\theta(y))$. 
We note that considering $\theta = \theta(y)$ in Eq.~\ref{eq_J_Cross} as a function of the impact parameter and taking into account that $d\sigma_s/d\theta = dy(\theta)/d\theta$ one can rewrite the formula for $\mathcal{J}_{ss}$ in the following~way:
\begin{equation}
\label{eq_J_Class}
\mathcal{J}_{ss} = \frac{1}{2\pi \hbar} \int\limits_{-\infty}^{\infty} p_y^s(y) dy, 
\end{equation}
which is now expressed only 
in terms of 
the transverse momentum projection $p_y^s(y)$ resulting from the classical trajectory with the impact parameter $y$. 

The next step is to determine $p_y^s(y)$ which an electron gains moving inside a chiral potential. 
The condition of the adiabaticity indicates that the electron spin quantization axis rotates in space 
following the local direction of the magnetization $\bm{n}(\bm{r})$. 
During this process the electron wavefunction acquires 
a geometrical phase, which is the Berry phase when considered along a closed loop~\cite{berry1984quantal}. 
Essentially, it also manifests itself as the appearance of an effective magnetic field acting on the electron orbital motion~\cite{Ye1999,BrunoDugaev,sundaram1999wave}.
This effect finds a classical explanation, which is described in details by Aharonov and Stern~\cite{AharonovStern}. 
The equation of motion for an electron experiencing the 
adiabatic rotation of its spin axis is~given:
\begin{equation}
\frac{d \boldsymbol{p}_s}{dt} = \frac{|e|}{c} \left[ \boldsymbol{v} \times \boldsymbol{B}^{s}(\boldsymbol{r}) \right] 
\label{eq_Newton}
\end{equation}
here 
$B_z^{s}$ is an effective spin-dependent magnetic field~\cite{BrunoDugaev} which is responsible for the appearance of the topological Hall effect; 
it can be related to the geometrical characteristic of the magnetization field, namely:
\begin{equation}
\label{eq_Bezz}
B_z^{\pm 1/2}(\bm{r}) = \pm \phi_0 \rho_{sk}(\bm{r}),
\quad 
\rho_{sk}(\bm{r}) = \frac{1}{4\pi} \boldsymbol{n}(\boldsymbol{r}) \cdot \left[ \partial_x \boldsymbol{n}(\boldsymbol{r}) \times \partial_y \boldsymbol{n}(\boldsymbol{r}) \right].
\end{equation}
where 
$\phi_0 = hc/|e|$ is the magnetic flux quantum, and 
$\rho_{sk}(\bm{r})$ is the skyrmion density; 
we note that the unit vector in Eq.~\ref{eq_Bezz} obeys $\bm{n}_z(r\to \infty) = \bm{e}_z$, so the sign of $\rho_{sk}$ is merely determined by $\chi$. 
The total integral over $\rho_{sk}$ gives the topological charge of a spin texture:
\begin{equation}
Q = \int \rho_{sk}(\bm{r}) d\bm{r} = 0, \pm 1, \pm 2 \dots,
\end{equation}
the configurations having $Q\neq 0$ are classified as magnetic skyrmions~\cite{NagaosaNature}. 
Using Eq.~\ref{eq_Newton} we express 
$p_y^s(y)$ in the following way:
\begin{equation}
\label{eq_IMP}
p_y^s(y) = -\frac{|e|}{c} \int v_x(t) B_z^{s}(x(t),y(t)) dt, 
\end{equation}
where the integration goes over the time of the electron presence inside a scattering region, and $(x(t),y(t))$ is its classical trajectory. 
We further assume 
that the scattering has a small-angle character (it is typical for large scale potentials $kr_0 \gg 1$),
i.e. the obtained transverse momentum 
$p_y^s \ll p_x^s$ 
is small compared to its initial value.
At that one can further simplify the integration in Eq.~\ref{eq_IMP} by replacing
the coordinate $y(t)$ of 
the real trajectory by its initial position $y(t) \approx y$ and by imposing the integration 
 $v_x(t) dt \approx dx$ 
over the straight line. 
The resulting expression for $p_y^s(y)$ is given:
\begin{equation}
p_y^s(y) =
-\frac{|e|}{c} \int B_z^{s}(x(t),y) v_x(t) dt \approx 
-\frac{|e|}{c} \int\limits_{-\infty}^{\infty} B_z^{s}(x,y) dx. 
\end{equation}
Finally, we substitute this formula into Eq.~\ref{eq_J_Class} 
and get for the total asymmetric rates:
\begin{equation}
\mathcal{J}_{\downarrow \downarrow}
= - \mathcal{J}_{\uparrow \uparrow} = 
\int \rho_{sk}(\bm{r})d\bm{r} = Q.
\label{eq_QTOP}
\end{equation}
Thus we obtained a remarkable finding, 
namely $\mathcal{J}_{ss}$ 
are entirely determined by the topological charge $Q$ of a spin texture. 
The magnitude of the transverse current appears to be robust and independent of a particular distribution of magnetization inside a skyrmion core. 

Let us emphasize, however, 
that the topological "quantization" does not have a universal character. 
Indeed, the result from Eq.~\ref{eq_QTOP} remains valid only
upon three additional assumptions, 
namely the classical character of an electron motion, the small-angle character of the scattering and the adiabaticity of the electron spin motion. 
Apart from these assumptions 
the topology of a spin texture ceases to be the unique requirement for the appearance of the scattering asymmetry.  In particular, the expressions~\ref{eq_ab}, \ref{eq_J_cons1} for $\mathcal{J}_{\bm{kk}'}^{ss'}$ obtained via the perturbation theory do not reflect any topological features of a scattering potential; the asymmetric scattering takes place independently of $Q$ in that case.

It is worth mentioning that the sign of the effective magnetic field $B_z^s$ is opposite for two electron spin state, so the resulting asymmetry $\mathcal{J}_{\uparrow \uparrow} = - \mathcal{J}_{\downarrow \downarrow}$ 
in the adiabatic regime 
is spin-dependent which naturally leads to the spin Hall effect. 
This is in contrast to the behavior of the electron scattering on a pure spin texture in the weak coupling regime found in Eq.~\ref{eq_J_cons1}, when the scattering asymmetry is spin-independent. 

The present consideration is equally applicable to the single subband case ($\Delta>2E$). 
At sufficiently large potential radius 
the condition of adiabaticity becomes fulfilled as well, at that the Berry phase approach is also valid. 
Therefore we argue that the magnitude of the total asymmetric rate for the single subband regime will be also determined by Eq.~\ref{eq_QTOP}. 



\section{Numerical solution (Methods)}

\label{s4}

In this section we present the scheme for the numerical calculations of $T$-matrix describing 
the electron scattering on potentials from  Eq.~\ref{eq_Vchiral}. 
The exact solution of the scattering problem is especially important 
for the investigation of 
the spin motion crossover which occurs 
when one passes from a perturbative scattering 
to the classical motion. 
The section has mostly a methodological character as it provides an alternative platform for the numerical studies of THE. 
The readers mainly interested in 
physical properties of the scattering can go straight to~\ref{s5}, where 
the results obtained by the numerical calculations are discussed in details. 



The considered potentials ${V}(\boldsymbol{r})$ have an important feature, namely 
the following commutator turns out to be zero:
\begin{equation}
\left[ V(\bm{r}), - i \partial_\phi + \chi \hat{\sigma}_z/2 \right] =0. 
\end{equation}
The operator $ - i \partial_\phi + \chi \hat{\sigma}_z/2 $ has the meaning of $z$-component of the total angular momentum.
The existence of such an integral of motion allows us to  separate the polar coordinates $\bm{r}=(r,\phi)$ in the Schrodinger equation, which opens up a way towards the application of the phase theory of scattering. 
However, a specific angular structure of the associated eigenfunctions modifies the decomposition of $T$-matrix on its partial scattering parameters. 
The further consideration goes as follows.
Firstly in~\ref{sDec} 
we derive the expansion of $T$-matrix in terms of the eigenstates associated with  $ - i \partial_\phi + \chi \hat{\sigma}_z/2 $, 
and secondly in~\ref{s_PFM}
we adjust the phase-function method for the numerical calculations of the scattering parameters. 
The analysis given in sections~\ref{s4},\ref{s5} is applicable for the case $E>\Delta/2$ when two spin subbands are activated. 
The single subband regime is considered separately in~\ref{sOne}.



\subsection{Angular harmonics}

\label{sEig}

The angular harmonics $\psi_m(\bm{r})$ corresponding to the operator $ - i \partial_\phi + \chi \hat{\sigma}_z/2 $ are given by:
\begin{equation}
\psi_m(r,\phi) = e^{i m \phi} \begin{pmatrix}
a_m(r)\\
e^{i(\chi \phi + \gamma)} b_m(r)
\end{pmatrix}, 
\label{eq_psim}
\end{equation}
here $m=0, \pm 1, \pm 2, \dots$ takes integer values, the functions $a_m,b_m$ depend only on $r$. One can naturally see that $\psi_m$ are the eigenfunctions of $-i\partial_{\phi} + \chi \sigma_z/2$ 
with the 
eigenvalue $m+\chi/2$. 
The functions $a_m(r),b_m(r)$ are determined by the explicit form of $v_{1,2}(r),u(r)$. 
The two-component functions $g_m \equiv (a_m,b_m)^T$ satisfy the following matrix equation: 
\begin{align}
&
\hat{\mathcal{H}}_m
g_m(r)
= - \omega_0 \hat{W}(r)
g_m(r),
\label{eq_ShMatrix}
\\
&
\hat{\mathcal{H}}_m = 
\begin{pmatrix}
\frac{1}{r} \partial_r r \partial_r  - \frac{m^2}{r^2} + k_{\uparrow}^2 & 0
\\
0  & \frac{1}{r} \partial_r r \partial_r  - \frac{(m+\chi)^2}{r^2} + k_{\downarrow}^2
\end{pmatrix} ,
\hspace{0.5cm}
\hat{W}(r) = \begin{pmatrix}
v_{1}(r) & u(r)
\\
u(r)  & v_{2}(r)	
\end{pmatrix},
\notag
\end{align}
where $\omega_0={2 m_{0} g}/{\hbar^2} $ and we assume $E>\Delta/2$. 
Solving Eq.~\ref{eq_ShMatrix} gives us the relevant scattering parameters needed for the computation of $T$-matrix. 

\subsection{Decomposition of $T$-matrix}

\label{sDec}


We note that the left part of Eq.~\ref{eq_ShMatrix} describes an electron free motion, so that 
away from the scattering potential $r>r_{0}$ the rigth side of Eq.~\ref{eq_ShMatrix} is absent and 
there are two independent cylindrical waves 
$g_m^{1,2}(r)$ given by: 
\begin{align}
& g_m^1 = \begin{pmatrix}
J_m(k_{\uparrow}r)
- \mathcal{K}_m^{11} Y_m(k_{\uparrow}r) 
\\
-  \mathcal{K}_m^{21}
Y_{m+\chi}(k_{\downarrow}r) 
\end{pmatrix},
\hspace{0.2cm}
& g_m^2 = \begin{pmatrix}
- \mathcal{K}_m^{12}
Y_{m}(k_{\uparrow}r) 
\\
J_{m+\chi}(k_{\downarrow}r)
-  \mathcal{K}_m^{22} Y_{m+\chi}(k_{\downarrow}r)
\end{pmatrix},
\label{eq_g12}
\end{align}
where $J_m, Y_m$ are the Bessel functions of the first and second kind respectively, the matrices $\hat{\mathcal{K}}_m$ of constant coefficients $\mathcal{K}_m^{ij}$ ($i,j=1,2$) are determined by $\hat{W}(r)$ profile at $r<r_0$. 
Our goal is to express $T$-matrix through $\hat{\mathcal{K}}_m$ coefficients.

Let us consider a wave function $\Psi(\boldsymbol{r})$  which satisfies the full Eq.~\ref{eq_First} with energy $E$ and which has the following asymptotic form away from the scattering potential $r \gg r_0$:
\begin{align}
& \Psi({r},\varphi) =
 \psi_{in} + \psi_{sc}, 
 \notag\\
& \psi_{in} = \begin{pmatrix}
e^{i {\boldsymbol{k}_{\uparrow}'} \boldsymbol{r}} u_{\uparrow} 
\\
e^{i {\boldsymbol{k}_{\downarrow}'}\boldsymbol{r}} u_{\downarrow} 
\end{pmatrix}, 
\hspace{0.5cm}
\psi_{sc}(r,\varphi)= \frac{1}{\sqrt{r}} \begin{pmatrix}
e^{i k_{\uparrow} r} \left( f_{\uparrow \uparrow} u_{\uparrow}  + f_{\uparrow \downarrow} u_{\downarrow}  \right)
\\
e^{i k_{\downarrow} r} \left( f_{\downarrow \uparrow} u_{\uparrow}  + f_{\downarrow \downarrow} u_{\downarrow}  \right)
\end{pmatrix},
\label{eq_WF1}
\end{align}
here 
$\bm{r}=(r,\varphi)$ is the radius vector in the polar coordinates, the function $\psi_{in}$ describes the incident plane wave with energy $E$ and momentum direction  $\boldsymbol{n}' = (\cos{\varphi'},\sin{\varphi'})$, the magnitude of the wavevector differs for two spin subband $\bm{k}_s' = (k_s',\varphi')$, where $k_s'$ are given in Eq.~\ref{eq_wvNom}, the polar angle $\varphi'$ corresponds to the direction on the incident flux; 
the coefficients $u_{1,2}$ determine the incident spin polarization of the electron ($|u_1|^2 + |u_2|^2 =1$).
The second term 
$\psi_{sc}$ corresponds to the outgoing cylindrical scattered wave, 
$f_{ss'}(\varphi,\varphi')$ is the scattering amplitude; here $\varphi$ is regarded as the polar angle of the scattered plane wave described by the wavevectors $\bm{k}_s = (k_s,\varphi)$ so that the scattering angle is defined as $\theta = \varphi - \varphi'$. 
The scattering amplitude $f_{ss'}(\varphi,\varphi')$ is connected with $T$-matrix at the mass shell as~\cite{adhikari1986quantum}:
\begin{equation}
\label{eq_Tf}
T_{\bm{kk}'}^{ss'} = - \frac{\hbar^2}{m_0} \sqrt{\frac{2\pi k_s}{i}} f_{ss'}(\varphi,\varphi'). 
\end{equation}

We further decompose $\Psi(\boldsymbol{r})$ over the set of the angular  harmonics $\psi_m$ from Eq.~\ref{eq_psim}: 
\begin{align}
&\Psi(r,\varphi) = 
 \sum_m i^m e^{-i m \varphi'} \left( \mathcal{A}_m^1 \psi_m^1(\bm{r}) + \mathcal{A}_m^2 \psi_m^2(\bm{r}) \right),
\end{align}
where $\mathcal{A}_m^{1,2}$ are some coefficients, and the functions $\psi_m^{1,2}$ taken at $r>r_0$ correspond to two linearly independent solutions $g_m^{1,2}(r)$ given by Eq.~\ref{eq_g12}. 
The divergent and convergent parts of the full $\Psi = \Psi^+ + \Psi^-$ and the incident $\psi_{in} = \psi_{in}^+ + \psi_{in}^-$ wave functions at $r>r_0$ are given~by:
\begin{align}
& \psi^\pm = \frac{1}{2}
\sum_m  i^m  e^{i m \theta}\begin{pmatrix}
u_1 H_m^\pm(k_{\uparrow}r)
\\
u_2 H_m^\pm(k_{\downarrow}r)
\end{pmatrix}, 
\label{eq_HankelWaves}
\\
& \Psi^\pm = \frac{1}{2} \sum_m i^m e^{i m \theta} \left[ \mathcal{A}_m^1 \begin{pmatrix}
\left(1 \pm i \mathcal{K}_m^{11} \right) H_m^\pm(k_{\uparrow}r) \\
e^{i(\chi \theta+\bar{\gamma}) } (\pm i \mathcal{K}_m^{21}) H_{m+\chi}^\pm(k_{\downarrow}r) 
\end{pmatrix}
+ \mathcal{A}_m^2 \begin{pmatrix}
(\pm i \mathcal{K}_m^{12}) H_{m}^\pm(k_{\uparrow}r) 
\\
e^{i(\chi \theta+\bar{\gamma})} \left(1 \pm i \mathcal{K}_m^{22} \right)H_{m+\chi}^\pm(k_{\downarrow}r) 
\end{pmatrix}
\right], 
\notag
\end{align}
where 
$\bar{\gamma}=\gamma + \chi \varphi'$, and $H_m^{\pm}$ are 
the Hankel functions of the first and second kind respectively. 
The scattered wave $\psi_{sc} = \Psi - \psi_{in}$ does not contain a convergent part, which 
brings us to the following system of equations on $\mathcal{A}_m^{1,2}$:
\begin{align}
& 
\begin{pmatrix}
1 - i \mathcal{K}_m^{11} & - i \mathcal{K}_m^{12}  \\
- i \mathcal{K}_m^{21} & 1 - i \mathcal{K}_m^{22} 
\end{pmatrix}
\begin{pmatrix}
\mathcal{A}_m^{1}\\
\mathcal{A}_m^{2}
\end{pmatrix}
= 
\begin{pmatrix}
u_1 \\
u_2 e^{i \delta}
\end{pmatrix},
\qquad \delta = \pi \chi/2 - \bar{\gamma}.
\end{align}
The solutions of these equations are given by: $(\mathcal{A}_m^1,\mathcal{A}_m^2)^T = \left(\hat{I} - i \hat{\mathcal{K}}_m\right)^{-1} (u_1, u_2 e^{i\delta})^T$,
where $\hat{I}$ is the unit matrix $2\times2$. 
Let us mention the appearance of an additional phase factor~$\delta$. 
The comparison of the scattered wave $ \psi_{sc} = \Psi^+ - \psi_{in}^+$ in the asymptotic region for the Hankel function $ H_m^+(x) \to (-i)^m e^{ix} \sqrt{(2/i\pi x)} $ 
with the expression for $\psi_{sc}$ containing $f_{ss'}(\varphi,\varphi')$  
leads us to 
the following expression for the scattering amplitude $f_{ss'}(\varphi,\varphi')$:
\begin{align}
f_{ss'}(\varphi,\varphi') = \frac{1}{\sqrt{2\pi i}} \sum_m e^{i m \theta}  \begin{pmatrix}
\frac{1}{\sqrt{k_{\uparrow}}} \left(S_m^{11} -1 \right)
& \frac{1}{\sqrt{k_{\uparrow}}}   S_m^{12} e^{-i(\chi \varphi' + \gamma')}
\\
e^{i(\chi \varphi + \gamma')}\frac{1 }{\sqrt{k_{\downarrow}}} S_m^{21} 
&e^{i\chi \theta} \frac{1}{\sqrt{k_{\downarrow}}} (S_m^{22} -1)
\end{pmatrix}_{ss'},
\label{eq_F-Sm}
\end{align}
where $\gamma' = \gamma - \pi \chi/2$ and 
we introduced the partial $\hat{S}_m$-matrices according to:
\begin{equation}
\label{eq_F-Sm}
\hat{S}_m = \left(\hat{I} + i \hat{\mathcal{K}}_m\right) \cdot \left(\hat{I} - i \hat{\mathcal{K}}_m\right)^{-1}. 
\end{equation} 
The present coupling between $\hat{S}_m$ and $\hat{\mathcal{K}}_m$ is commonly known for multichannel scattering problems.  
Using the relation \ref{eq_Tf} we finally get the decomposion of $T$-matrix:
\begin{equation}
T_{\boldsymbol{kk}'}^{ss'}
= \frac{{i}}{2\pi \nu_0} \sum_m e^{i m \theta}  \begin{pmatrix}
S_m^{11} -1 
&S_m^{12} e^{-i (\chi \varphi' + \gamma')}
\\
S_m^{21} e^{i(\chi \varphi + \gamma')}
&e^{i\chi \theta} (S_m^{22} -1)
\end{pmatrix}_{ss'},
\label{eq_Tfss}
\end{equation}
where the coefficients $S_m^{ij}$ are determined by a particular spatial profile $\hat{W}(r)$. 

\subsection{Phase-function method}

\label{s_PFM}

In this section we describe the numerical method for the calculation of $\hat{S}_m$, $\hat{\mathcal{K}}_m$ parameters entering in Eq.~\ref{eq_Tfss}, \ref{eq_F-Sm}. 
The Shrodinger equation is of the second order thus it requires two boundary conditions. 
In order to eliminate the necessity to address the wave function asymptotics at $r \gg r_0$ 
one uses the so-called phase function method~\cite{Babikov-book,Babikov1}, which replaces the second order Eq.~\ref{eq_ShMatrix} by the first order nonlinear Cauchy problem for a set of scattering parameters. The Cauchy problem can be further solved using the standard computational software. 
Here we provide step by step derivation of this method purposely, 
so that one could straightforwardly adjust the similar calculations for more complex band structures. 


Let us write Eq.~\ref{eq_ShMatrix} in the following form:
\begin{align}
\label{eq_SHR-phaseFunc}
&
\Bigl(
\hat{\mathcal{H}}_m^0 + \hat{I} r^{-1} \partial_r r \partial_r
\Bigr) g_m(r) 
=  - \omega_0 \hat{W}(r) g_m(r), 
\end{align}
where $
\hat{\mathcal{H}}_m^0 = {\rm diag}\left(k_{\uparrow}^2-m^2/r^2, k_{\downarrow}^2-(m+\chi)^2/r^2 \right)
$ and $\hat{I}$ is the unit matrix $2\times 2$. 
Following the textbook~\cite{Babikov-book,Babikov1} 
for a multichannel scattering we present the 
functions $g_m(r)$ as:
\begin{align}
\label{eq_WF}
&g_m(r) = \left( \hat{J}_m(r) - \hat{Y}_m(r) \hat{\mathcal{K}}_m(r) \right) \mathcal{C}_m(r),
\\
&\hat{J}_m(r) = \begin{pmatrix}
J_m(k_{\uparrow}r) & 0
\\
0 & J_{m+\chi}(k_{\downarrow}r) 
\end{pmatrix},
\hspace{0.2cm}
\hat{Y}_m(r) = \begin{pmatrix}
Y_m(k_{\uparrow}r) & 0
\\
0 & Y_{m+\chi}(k_{\downarrow}r) 
\end{pmatrix},
\notag
\end{align}
where the $2\times2$ matrix $\hat{\mathcal{K}}_m(r)$ and the two-component column $\mathcal{C}_m(r)$ are some functions of the coordinate $r$.  
Outside the scattering region $r>r_0$ the functions $g_m(r)$ can be present in form~\ref{eq_g12} with the matrix of $r$-independent coefficients $\hat{\mathcal{K}}_m$;
at that the normalization column $\mathcal{C}_m$ will describe the polarization structure of an electron state. 
In order to endow $\hat{\mathcal{K}}_m(r)$ with 
the meaning of the real scattering parameters on the potential 
$\hat{W}(\tilde{r}) \theta(r-\tilde{r})$ cut off at point $r<r_0$ one has to impose the additional condition for the derivative of~$g_m$:
\begin{align}
\frac{dg_m}{dr} = \left( \frac{d \hat{J}_m}{dr} - \frac{d \hat{Y}_m}{dr} \hat{\mathcal{K}}_m(r) \right) \mathcal{C}_m(r).
\label{eq_Der1PhaseFunc} 
\end{align}
The matrices $\hat{\mathcal{K}}_m(r)$ 
satisfying both Eq.~\ref{eq_SHR-phaseFunc},\ref{eq_Der1PhaseFunc} 
and taken at the boundary point $\hat{\mathcal{K}}_m(r_0)$ will correspond to the real scattering parameters of a potential $\hat{W}(r)$. 
The introduced functions $\hat{\mathcal{K}}_m(r)$ are called the phase functions.

We further proceed with the derivation of the first order equation on $\hat{\mathcal{K}}_m(r)$. 
Let us substitute $g_m(r)$ in Eq.~\ref{eq_SHR-phaseFunc} and take into account the condition \ref{eq_Der1PhaseFunc}. The term containing only the second derivatives of $g_m$ can be written as:
\begin{align}
\frac{1}{r} \frac{d}{dr} r \frac{d}{dr}  \left( \hat{J}_m - \hat{Y}_m \hat{\mathcal{K}}_m \right) \mathcal{C}_m = \left[\left( \hat{J}_m'' + \hat{J}_m'/r \right) - \left( \hat{Y}_m'' + \hat{Y}_m'/r \right) \hat{\mathcal{K}}_m\right] \mathcal{C}_m+ 
\notag
\\
+ \left( \hat{J}_m' - \hat{Y}_m' \hat{\mathcal{K}}_m \right) \frac{d \mathcal{C}_m}{dr} - \hat{Y}_m' \frac{d \hat{\mathcal{K}}_m}{dr} \mathcal{C}_m.
\end{align}
The terms in the square brackets from above cancel out $\hat{\mathcal{H}}_m^0 g_m$ term in Eq.~\ref{eq_SHR-phaseFunc}, 
thus Eq.~\ref{eq_SHR-phaseFunc} does not contain the second derivatives: 
\begin{equation}
\left( \hat{J}_m' - \hat{Y}_m' \hat{\mathcal{K}}_m \right)  \frac{d \mathcal{C}_m}{dr}  - \hat{Y}_m' \frac{d \hat{\mathcal{K}}_m }{dr} \mathcal{C}_m = - \omega_0 \hat{W}(r) \left( \hat{J}_m - \hat{Y}_m \hat{\mathcal{K}}_m  \right)\mathcal{C}_m.
\label{eq_Shr2-PhaseFunc}
\end{equation}
The next step is to express $d\mathcal{C}_m/dr$ through $d \hat{\mathcal{K}}_m/dr$. 
After multiplying this formula on $(\hat{J}_m - \hat{\mathcal{K}}_m \hat{Y}_m)$ 
we get for the left side of the equation:
\begin{align}
& \left( \hat{J}_m' - \hat{\mathcal{K}}_m \hat{Y}_m'  \right) \left( \hat{J}_m - \hat{Y}_m \hat{\mathcal{K}}_m \right) \frac{d \mathcal{C}_m}{dr} - \left( \hat{J}_m -  \hat{\mathcal{K}}_m \hat{Y}_m \right) \hat{Y}_m' \frac{d \hat{\mathcal{K}}_m}{dr} \mathcal{C}_m + \left[ \hat{\mathcal{K}}_m , \hat{\mathcal{W}}_m \right] \frac{d \mathcal{C}_m}{dr}, 
\label{eq_PhaseFunc-2}
\end{align}
where $\hat{\mathcal{W}}_m  = \hat{Y}_m' \hat{J}_m -\hat{Y}_m \hat{J}_m'$ is the Wronskian matrix of the Bessel functions and  $\left[ \hat{\mathcal{K}}_m , \hat{\mathcal{W}}_m \right] $ is the commutator of two matrices. 
Using the condition ~\ref{eq_Der1PhaseFunc} one can further express $ (\hat{J}_m - \hat{Y}_m \hat{\mathcal{K}}_m) \cdot d \mathcal{C}_m/dr$ through $d \hat{\mathcal{K}}_m/ dr$,
so Eq.~\ref{eq_SHR-phaseFunc} takes form: 
\begin{align}
- \hat{\mathcal{W}}_m \frac{d \hat{\mathcal{K}}_m}{dr} \mathcal{C}_m + \left[ \hat{\mathcal{K}}_m , \hat{\mathcal{W}}_m \right]  \frac{d \mathcal{C}_m}{dr} = - \omega_0 \left( \hat{J}_m - \hat{\mathcal{K}}_m \hat{Y}_m  \right) \hat{W}(r)  \left( \hat{J}_m - \hat{Y}_m \hat{\mathcal{K}}_m  \right)\mathcal{C}_m.
\end{align}
Noting that 
$\left[ \hat{\mathcal{K}}_m , \hat{\mathcal{W}}_m \right] =0$ due to 
 $\hat{\mathcal{W}}_m = \hat{I} \times (2/\pi r)$ we  
finally get the equation only in terms of $\hat{\mathcal{K}}_m(r)$ functions:
\begin{equation}
\label{eq_K-Phase}
\frac{d \hat{\mathcal{K}}_m}{dr} = \frac{\pi r}{2} \omega_0  \left(\hat{J}_m - \hat{\mathcal{K}}_m \hat{Y}_m\right) \hat{W}(r)  \left( \hat{J}_m - \hat{Y}_m \hat{\mathcal{K}}_m  \right).
\end{equation}
The initial condition is $\hat{\mathcal{K}}_m(0)=0$. 
The scattering matrices $\hat{S}_m$ can be further obtained by calculating this equation up to $r=r_0$ and using Eq.~\ref{eq_F-Sm}.

Alternatively, based on Eq.~\ref{eq_K-Phase},\ref{eq_F-Sm} one can derive the equation directly on $\hat{S}_m(r)$ functions. 
Introducing $S_m(r) \equiv (\hat{I} + i \hat{\mathcal{K}}_m(r))\cdot(\hat{I} - i \hat{\mathcal{K}}_m(r))^{-1}$, the derivative of $\hat{\mathcal{K}}_m$ can be expressed~as:
\[
i \frac{d\hat{\mathcal{K}}_m}{dr} = \frac{d\hat{S}_m}{dr}\left(\hat{I}-i \hat{\mathcal{K}}_m\right) - i \hat{S}_m \frac{d \hat{\mathcal{K}}_m}{dr}. 
\]
After troublesome algebra Eq.~\ref{eq_K-Phase} with the derivative from above is transformed to:
\begin{align}
\label{eq_S-Phase}
& \frac{d \hat{S}_m(r)}{dr} = i \frac{\pi r}{4} \omega_0 \left( \hat{H}_m^- + \hat{S}_m \hat{H}_m^+ \right) \hat{W}(r) \left( \hat{H}_m^- +  \hat{H}_m^+ \hat{S}_m \right), 
\\
& \hat{H}_m^\pm = \begin{pmatrix}
H_m^\pm(k_{\uparrow}r) & 0
\\
0 & H_{m+\chi}^\pm(k_{\downarrow}r) 
\end{pmatrix}, 
\qquad
\hat{W}(r) = \begin{pmatrix}
v_1(r) & u(r)
\\
u(r)  & v_2(r) 
\end{pmatrix}.
\notag
\end{align}
The initial condition is $\hat{S}_m(0) = \hat{I}$. 
We note, that Eq.~\ref{eq_S-Phase} is more convenient for numerical calcuations than Eq.~\ref{eq_K-Phase}. 
Indeed, in case of a single channel scattering problem the parameter $\mathcal{K}_m = \tan{\delta_m}$ would correspond to the tangent of the partial scattering phase $\delta_m$. 
When the potential radius is increased so are the scattering phases; particularly the latter can take value $|\delta_m| \to \pi/2$. At that $\mathcal{K}_m \to \infty$ diverges while $S_m = e^{2i \delta_m}$ remains finite. 
The analogous situation takes place for a multichannel scattering problem: when evaluating Eq.~\ref{eq_K-Phase} the functions $\hat{\mathcal{K}}_m(r)$ can turn to the infinity at some points, while Eq.~\ref{eq_S-Phase} and the functions $\hat{S}_m(r)$ are free of such negative feature. 


\section{Numerical solution (Results)}

\label{s5}

In this section we discuss the results for the scattering on chiral potentials 
obtained by the numerical calculations of Eq.~\ref{eq_S-Phase}.

\subsection{Scattering on a magnetic skyrmion in different regimes}

\label{sSkyrmion}

\begin{figure}[t]
	\centering	
	\includegraphics[width=1.\textwidth]{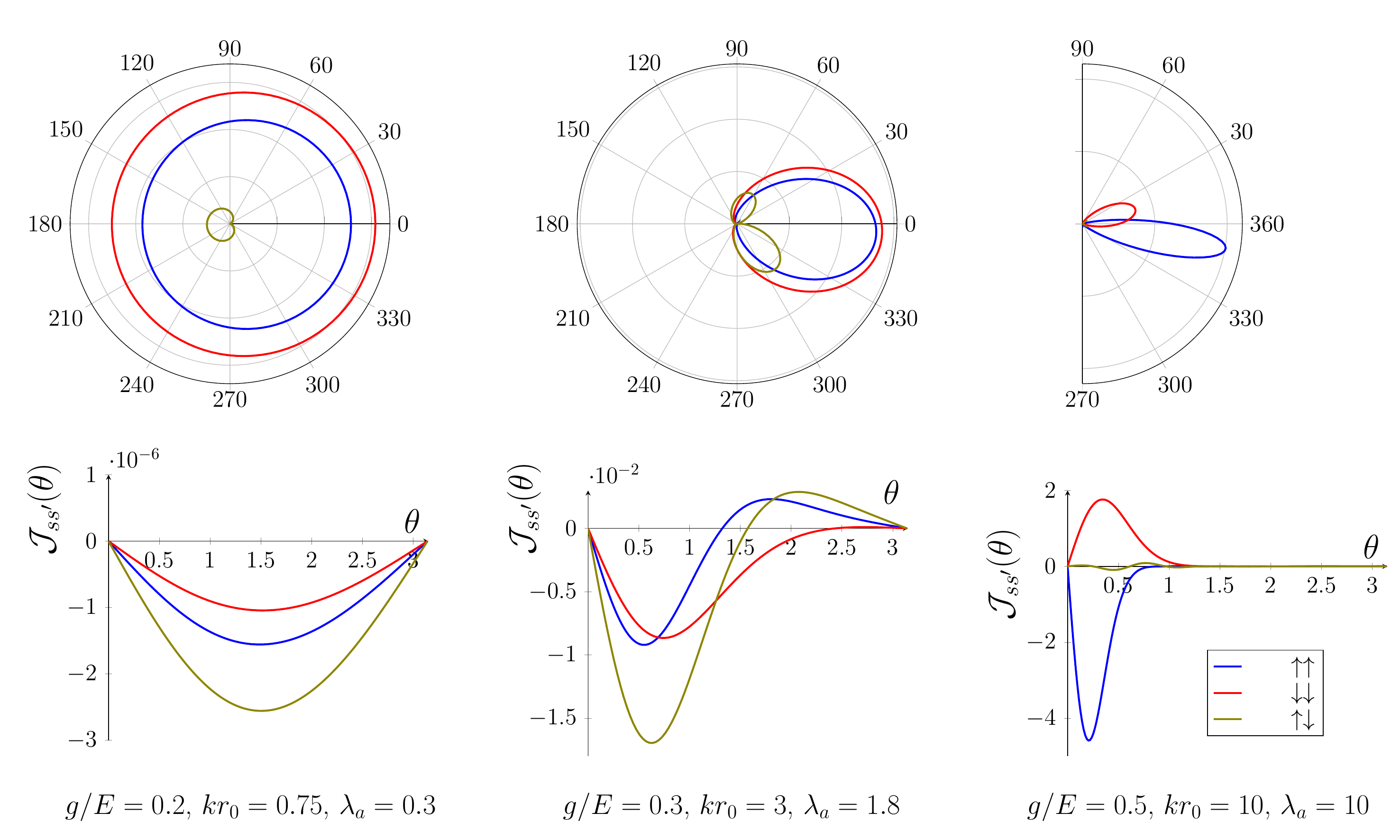}
	\caption{The scattering pattern and the asymmetric scattering rates $\mathcal{J}_{ss'}(\theta) \equiv \mathcal{J}_{\bm{kk}'}^{ss'}$. 
	}	
	\label{fpw}
\end{figure}

Here we consider the scattering on a magnetic skyrmion, namely we assume the following coupling between the scattering potential elements:
\begin{align}
\label{eq_Skyrmion_Not}
& v_1(r) = - v_2(r) = n_z(r)-1,
\quad u(r) = n_\parallel(r),
\qquad n_\parallel^2(r) + n_z^2(r) = 1.
\end{align}
We parametrize the functions 
$n_z = \cos{\Theta(r)}, n_\parallel = \sin{\Theta(r)}$
using the azimuthal angle of the magnetization field $\Theta(r)$. 
The spin splitting of the electron subbands $\Delta = 2g$. 
In the calculations shown below 
we take $\chi=1$ and make use of 
the following skyrmion profile: 
\begin{equation}
\label{eq_SkyrmParapmTheta}
\Theta(r) = \pi \left(1 - \frac{r}{r_0} \right),
\qquad r<r_0.
\end{equation}

In Fig.\ref{fpw} we demonstrate the computed scattering pattern along with the $\theta$-dependence of the asymmetric rates $\mathcal{J}_{ss'}(\theta) \equiv \mathcal{J}_{\bm{kk}'}^{ss'}$ for different scattering regimes. 
The left pannel corresponds to the weak-coupling regime.  
As we discussed in \ref{sWeak}, the scattering whithin the spin-conserving channels indeed has an isotropic-like character, while the spin-flip channels are characterized by the suppression of the forward scattering. 
The asymmetric rates show a $\sin$-like dependence on the scattering angle, 
the type of the scattering asymmetry is unique for all scattering channels, 
which is in full agreement with Eq.~\ref{eq_J_cons1},\ref{eq_J_WK_Charge}. 
The middle pannel  
demonstrates the crossover regime, here the scattering pattern starts narrowing into the forward direction and the asymmetric rates gradually lose a certain preferable direction. 
The right pannel shows the scattering in the adiabatic regime. 
The spin-flip scattering channels are suppressed in this case. The small-angle scattering taking place for the spin-conserving channels is featured by the pronounced spin-dependent asymmetry, 
the latter indicates the presence of the spin-dependent magnetic fields due to the Berry phase, see~\ref{sAdiab}. 

The symmetry crossover between 
the spin-dependent 
and the spin-independent 
Hall responses 
has been firstly discovered in~\cite{denisov-SciRep}; 
the detailed discussion of 
its features can be found in~\cite{denisov2018general}. 
Let us mention that to describe the behavior of the 
scattering rates in the crossover regime one 
necessarily has to address the exact solution of the scattering problem, at that the numerical scheme from \ref{s4} is of special importance. 

\subsection{Classical limit and the topology}

\label{s-Quasi}

In~\ref{sAdiab} we demonstrated that 
when both the classical and the adiabatic conditions are fulfilled 
($kr_0, \lambda_a\gg1$)
the total asymmetric rates $\mathcal{J}_{ss}$ are quantized with its magnitude determined by the topological charge $Q$ of a magnetization field. 
Below this effect is examined by the numerical calculations. 
We consider two purely magnetic potentials
featured by the different topology of parental spin textures. 
We take $\chi = 1$ and use the following spin texture profiles 
(the parametrization is introduced in Eq.~\ref{eq_Skyrmion_Not}, $\Delta = 2g$): 
\begin{align}
\label{eq_Q01}
& (Q=1) \qquad \Theta(r) = \pi \left(1 - \frac{r}{r_0} \right),
\qquad r<r_0;
\\
& (Q=0) \qquad \Theta(r) = 2 \pi \frac{r}{r_0} \left(1 - \frac{r}{r_0} \right),
\qquad r<r_0.
\notag
\end{align}

\begin{figure}[t]
	\centering	
	\includegraphics[width=0.5\textwidth]{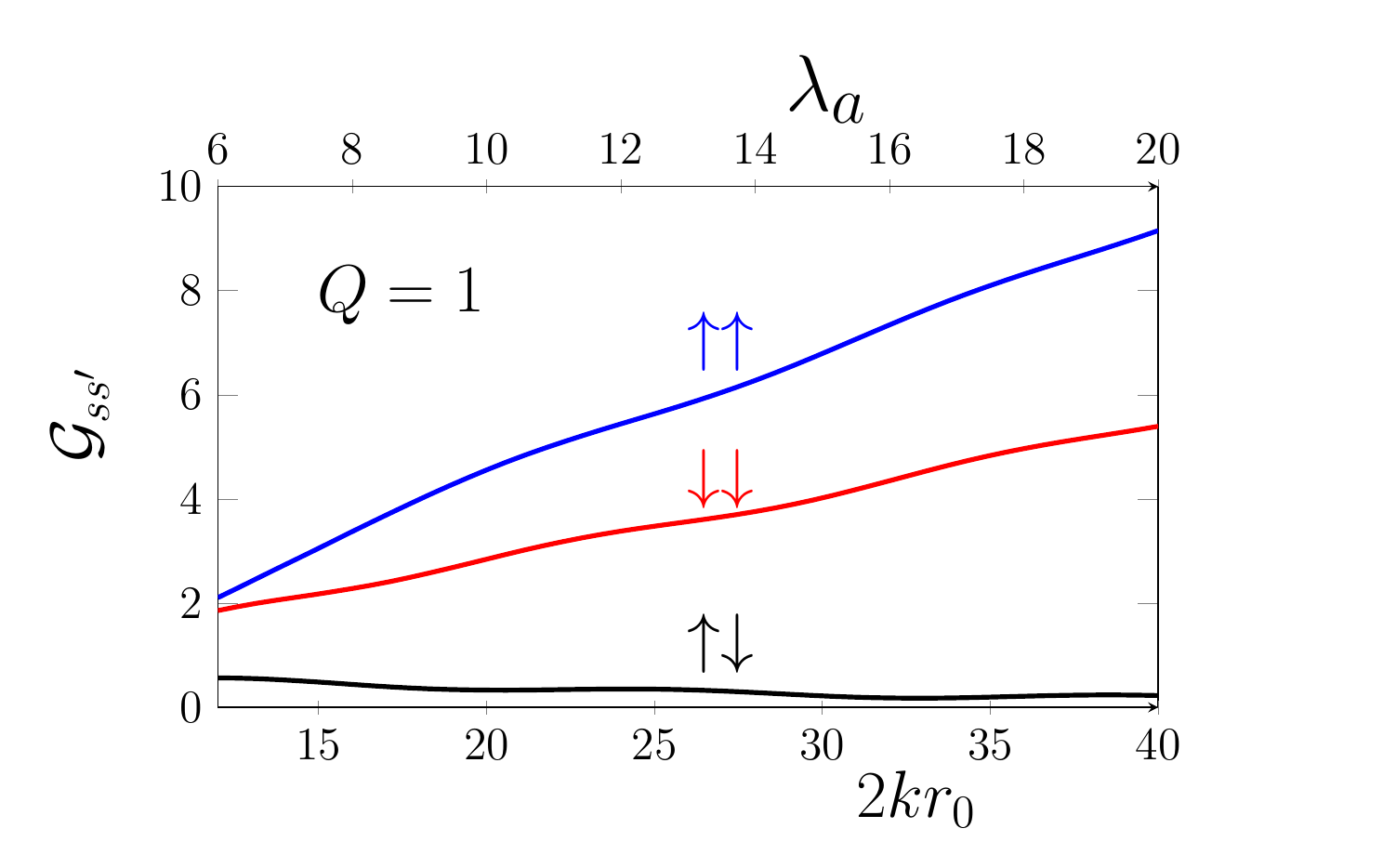}
	\includegraphics[width=0.5\textwidth]{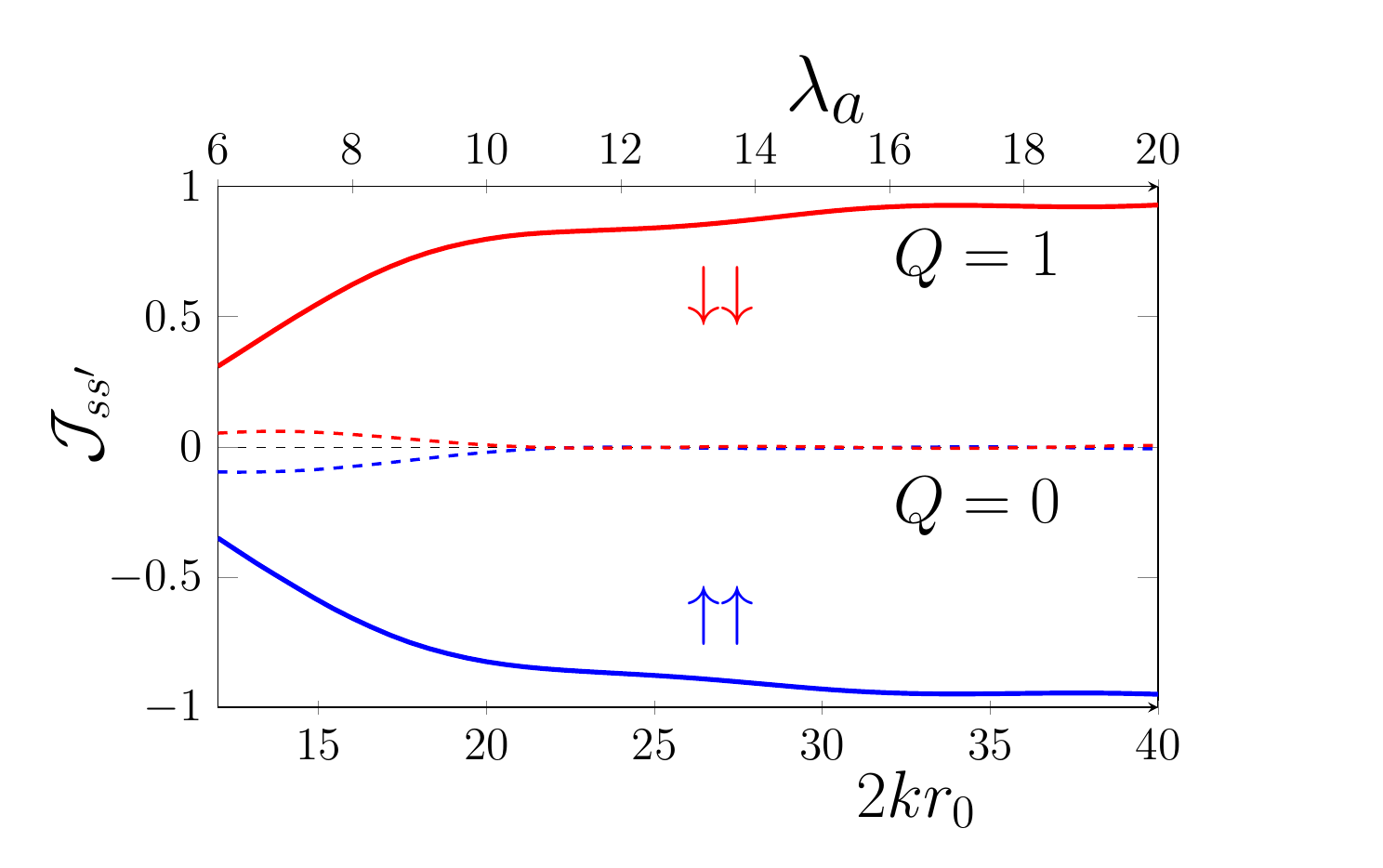}
	\caption{The dependence of the total symmetric $\mathcal{G}_{ss'}$ and asymmetric $\mathcal{J}_{ss'}$ rates on the potential radius in units $kr_0$ for $Q=0,1$ spin configurations,
		the ratio $g/E = 0.5$.
	}	
	\label{f3-7}
\end{figure}

In the right panel of Fig.~\ref{f3-7} we demonstrate the computed dependence of 
$\mathcal{J}_{ss}$ on the spin texture radius when entering into the classical and adiabatic limits ($g/E=0.5$ is fixed). 
This figure shows that 
there is a saturation of $\mathcal{J}_{ss}$ 
when $kr_0$ increases, 
the limiting magnitude of $|\mathcal{J}_{ss}|$ 
approaches $Q=1$ for the skyrmion configuration 
and goes down to zero for the topologically uncharged texture $Q=0$. 
This is a clear manifestation of the topological features 
discussed in~\ref{sAdiab}, namely the asymmetric rates 
approach the limiting values 
$\mathcal{J}_{\downarrow \downarrow} =-\mathcal{J}_{\uparrow \uparrow} =Q$ 
determined by the topology.

In the left panel of Fig.~\ref{f3-7} we demonstrate 
the behavior of the average symmetric rate $\mathcal{G}_{ss'}$
for different scattering channels in the same region of parameters as for $\mathcal{J}_{ss}$; 
here the data is present only for $Q=1$ configuration. 
We note that the magnitude of $\mathcal{G}_{\uparrow \downarrow}$ for the spin-flip channel indeed gets suppressed when $\lambda_a$ is increased, this is the consequence of the spin adiabaticity. 
It is worth mentioning that $\mathcal{G}_{ss}$ for the spin-conserving channels
increase linearly with the scatterer size, which is expected for a scattering in 2D systems. 
On the contrary, 
upon the increase of $r_0$ 
the asymmetric rates $\mathcal{J}_{ss}$ 
become independent in the region $kr_0 \gg 1$ 
not only on a scattering potential size, but on its particular inner structure as well.

\subsection{Scattering on electrically charged skyrmions}


\label{sCharge}

In this section we study 
the electron asymmetric scattering 
in case when a scalar potential is 
present additionally to a magnetic skyrmion. 
We make use of the following parametrization for the diagonal elements $v_{1,2}$:
\begin{align}
\label{eq_Pot-El}
& v_1(r) = n_z(r)-1 + \delta U(r), 
\qquad v_2(r) = 1- n_z(r) + \delta U(r), 
\end{align}
the off-diagonal component $u(r) = n_\parallel(r) = \sqrt{1-n_z^2(r)}$ remains determined by $n_z(r)$, see Eq.~\ref{eq_Skyrmion_Not}. The subband splitting $\Delta = 2g$.
For the numerical calculations we use the skyrmion profile from Eq.~\ref{eq_SkyrmParapmTheta} and the following form of the potential 
$\delta U(r) = U_0 e^{- (r/R)^2}$ with $R$ being its  localization radius. 

\begin{figure}[t]
	\centering	
	\includegraphics[width=1.\textwidth]{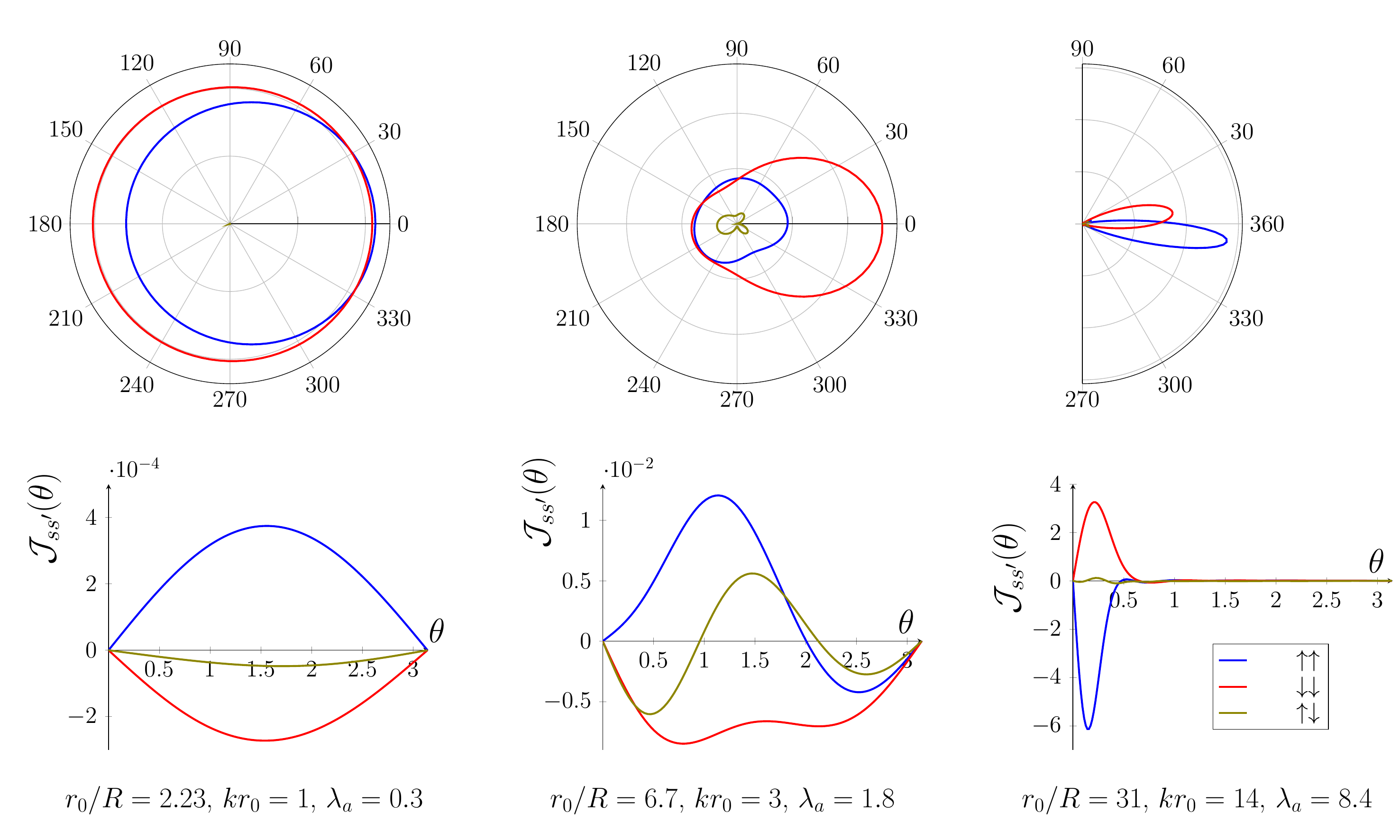}
	\caption{
The scattering pattern and the asymmetric scattering rates $\mathcal{J}_{ss'}(\theta) \equiv \mathcal{J}_{\bm{kk}'}^{ss'}$
in case of an additional short-range scalar potential. 
The parameters: 
$U_0 = 30$, $k R = 0.45$, $g/E = 0.3$. 
	}	
	\label{f-Electro}
\end{figure}

In Fig.~\ref{f-Electro} 
we demonstrate the scattering pattern and the $\theta$-dependence of $\mathcal{J}_{\bm{kk}'}^{ss'}$ for different skyrmion radii. 
In this plot only 
$r_0$ is varied, other parameters such as the localization radius $R$, scattering energy $E$ and $g,U_0$ remain unchanged.  
The left panel of Fig.~\ref{f-Electro} 
corresponds to the perturbative scattering regime, 
here the scalar potential and the skyrmion are close 
in size. 
It is seen from the left panel that $\mathcal{J}_{\bm{kk}'}^{\uparrow \uparrow},\mathcal{J}_{\bm{kk}'}^{\downarrow \downarrow}$
have different signs, 
which is consistent with the results 
of the weak coupling theory~\ref{sWeak} 
and Eq.~\ref{eq_J_WK_Spin}
predicting that the nonmagnetic component of ${V}(\bm{r})$ restores 
the spin Hall effect and suppresses the spin-independent scattering caused by a pure magnetic texture. 
The right pannel of Fig.~\ref{f-Electro} corresponds to the adiabatic regime with respect to the skyrmion size; here 
$\delta U(r)$ is kept localized ($r_0/R = 31$) 
so its length $R$ is small compared to $r_0$. 
The scattering rates in this regime have a similar structure to that in case of a pure magnetic skyrmion (see Fig.~\ref{fpw}). Therefore 
in case of a large skyrmion 
the short-range scalar perturbation does not affect significantly the asymmetric scattering, 
the latter is mainly produced during a lingering electron motion in the skyrmion texture surrounding $\delta U(r)$, at that
the general arguments given in~\ref{sAdiab} for the classical regime 
remain applicable. 
The middle panel in Fig.~\ref{f-Electro} corresponds to the intermediate case, here both the scattering pattern and $\mathcal{J}_{\bm{kk}'}^{ss'}$ are strongly influenced by the presence of $\delta U(r)$. 
We note that the scattering asymmetry in the weak coupling regime depends on the sign of $U_0$. 
Data shown in Fig.~\ref{f-Electro}  is obtained for the positive value $U_0 =30$, at that the scattering asymmetry for the spin up and the spin down states shown in the left panel differs from that in the classical limit. 
Increasing $r_0$ drives the scattering channels to switch their asymmetry, 
at that a complex scattering pattern at the intermediate region is indeed expected. 





\subsection{One spin subband regime}

\label{sOne}

In this section we consider the case when $E <\Delta/2$ and only the spin up subband 
is available for a free motion (see Fig.~\ref{f2}). 
Firstly we adjust the phase function method for this regime and secondly we discuss some numerical results. 

The following matrix equations on $g_m(r)$ functions should be used instead of Eq.~\ref{eq_ShMatrix}: 
\begin{align}
&
\hat{\mathcal{H}}_m'
g_m(r)
= - \omega_0 \hat{W}(r)
g_m(r),
\label{eq_ShMatrix2}
\\
&
\hat{\mathcal{H}}_m' = 
\begin{pmatrix}
\frac{1}{r} \partial_r r \partial_r  - \frac{m^2}{r^2} + k_\uparrow^2 & 0
\\
0  & \frac{1}{r} \partial_r r \partial_r  - \frac{(m+\chi)^2}{r^2} - \varkappa^2
\end{pmatrix} ,
\hspace{0.5cm}
\hat{W}(r) = \begin{pmatrix}
v_{1}(r) & u(r)
\\
u(r)  & v_{2}(r)	
\end{pmatrix},
\notag
\end{align}
where we introduced the real parameter 
$\varkappa =  \sqrt{2m_0(\Delta/2-E)}$. 
Two independent solutions $g_{m}^{1,2}$ 
of Eq.~\ref{eq_ShMatrix2}
away from the scattering potential $r>r_0$ 
are given by:
\begin{align}
& g_m^1 = \begin{pmatrix}
J_m(k_\uparrow r)
- \mathcal{K}_m^{11} Y_m(k_\uparrow r) 
\\
-  \mathcal{K}_m^{21}
K_{m+\chi}(\varkappa r) 
\end{pmatrix},
\hspace{0.2cm}
& g_m^2 = \begin{pmatrix}
- \mathcal{K}_m^{12}
Y_{m}(k_\uparrow r) 
\\
I_{m+\chi}(\varkappa r)
-  \mathcal{K}_m^{22} K_{m+\chi}(\varkappa r)
\end{pmatrix},
\label{eq_g34}
\end{align}
where $I_m,K_m$ are modified Bessel functions of the first and second kind respectively. 
Among all $\mathcal{K}_m^{ij}$ coefficients 
only $\mathcal{K}_m \equiv \mathcal{K}_m^{11}$ remains relevant for the scattering properties. 
Indeed, 
the asymptotic form of the propagating wavefunction at $r\gg r_0$ 
contains only spin-up state:
\begin{align}
\Psi'(r,\varphi) = \left(e^{i \boldsymbol{k' r}} + \frac{e^{ikr}}{\sqrt{r}} 
f(\varphi,\varphi') \right) |\uparrow \rangle .
\end{align}
Since $I_m$ entering in $|\downarrow \rangle$ state diverges at $r\gg r_0$
the expansion of $\Psi'$ over the partial harmonics $g_{m}^{1,2}$ cannot contain the admixture of 
$g_m^2$ functions. Therefore 
the scattering amplitude is given by a conventional single-channel decomposition:
\begin{equation}
\label{eq_f_2}
f(\varphi,\varphi') = \frac{1}{\sqrt{2\pi i k_\uparrow}} \sum_m e^{i m (\varphi-\varphi')} \left(\mathcal{S}_m -1\right), 
\hspace{0.5cm}
\mathcal{S}_m = \frac{1+i \mathcal{K}_m}{1-i \mathcal{K}_m}. 
\end{equation}

The phase-function method is adjusted for the one spin subband case using Eq.~\ref{eq_ShMatrix2},\ref{eq_g34}. 
We introduce the phase functions $\hat{\mathcal{K}}_m(r)$ according to the following notation:
\begin{align}
\label{eq_WF-2}
& g_m(r) = \left( \hat{Q}_m - \hat{Z}_m \hat{\mathcal{K}}_m \right) \mathcal{C}_m(r),
\hspace{0.5cm}
\frac{dg_m}{dr} \equiv \left( \frac{d\hat{Q}_m}{dr} - \frac{d\hat{Z}_m}{dr} \hat{\mathcal{K}}_m \right) \mathcal{C}_m(r),
\notag
\\
&\hat{Q}_m(r) = \begin{pmatrix}
J_m(k_{\uparrow}r) & 0
\\
0 & \sqrt{\frac{2}{\pi}} I_{m+\chi}(\varkappa r) 
\end{pmatrix},
\hspace{0.2cm}
\hat{Z}_m(r) = \begin{pmatrix}
Y_m(k_{\uparrow}r) & 0
\\
0 & - \sqrt{\frac{2}{\pi}} K_{m+\chi}(\varkappa r) 
\end{pmatrix},
\end{align}
where the normalization constant $\sqrt{2/\pi}$ is introduced to make the Wronskian $ \hat{Q}_m \hat{Z}_m' -  \hat{Q}_m' \hat{Z}_m = \hat{I} \times 2/\pi r$ proportional to the unity matrix. 
Using the functions $g_m$ from Eq.~\ref{eq_WF-2} and making the similar transformations of Eq.~\ref{eq_ShMatrix2} as were described in~\ref{s_PFM} 
we get the following equations for $\hat{\mathcal{K}}_m(r)$ and $\hat{S}_m(r)= (\hat{I} + i \hat{\mathcal{K}}_m)\cdot (\hat{I} + i \hat{\mathcal{K}}_m)^{-1}$ matrix functions:
\begin{align}
\label{eq_K-Phase2}
& \frac{d \hat{\mathcal{K}}_m}{dr} = \frac{\pi r}{2} \omega_0 \left(\hat{Q}_m - \hat{\mathcal{K}}_m \hat{Z}_m\right) \hat{W}(r)  \left( \hat{Q}_m - \hat{Z}_m \hat{\mathcal{K}}_m  \right),\\
& \frac{d\hat{S}_m}{dr} = 
i \frac{\pi r}{4} \omega_0 \left( \hat{Q}_m- i \hat{Z}_m + \hat{S}_m \cdot \left(\hat{Q}_m + i \hat{Z}_m\right) \right) \hat{W}(r) \left(  \hat{Q}_m- i \hat{Z}_m + \left(\hat{Q}_m + i \hat{Z}_m\right) \cdot \hat{S}_m \right). 
\notag
\end{align}
Let us mention that to calculate the scattering amplitude only $\mathcal{K}_m \equiv \mathcal{K}_m^{11}$ element of the whole matrix is needed.

\begin{figure}[t]
	\centering	
	\includegraphics[width=0.6\textwidth]{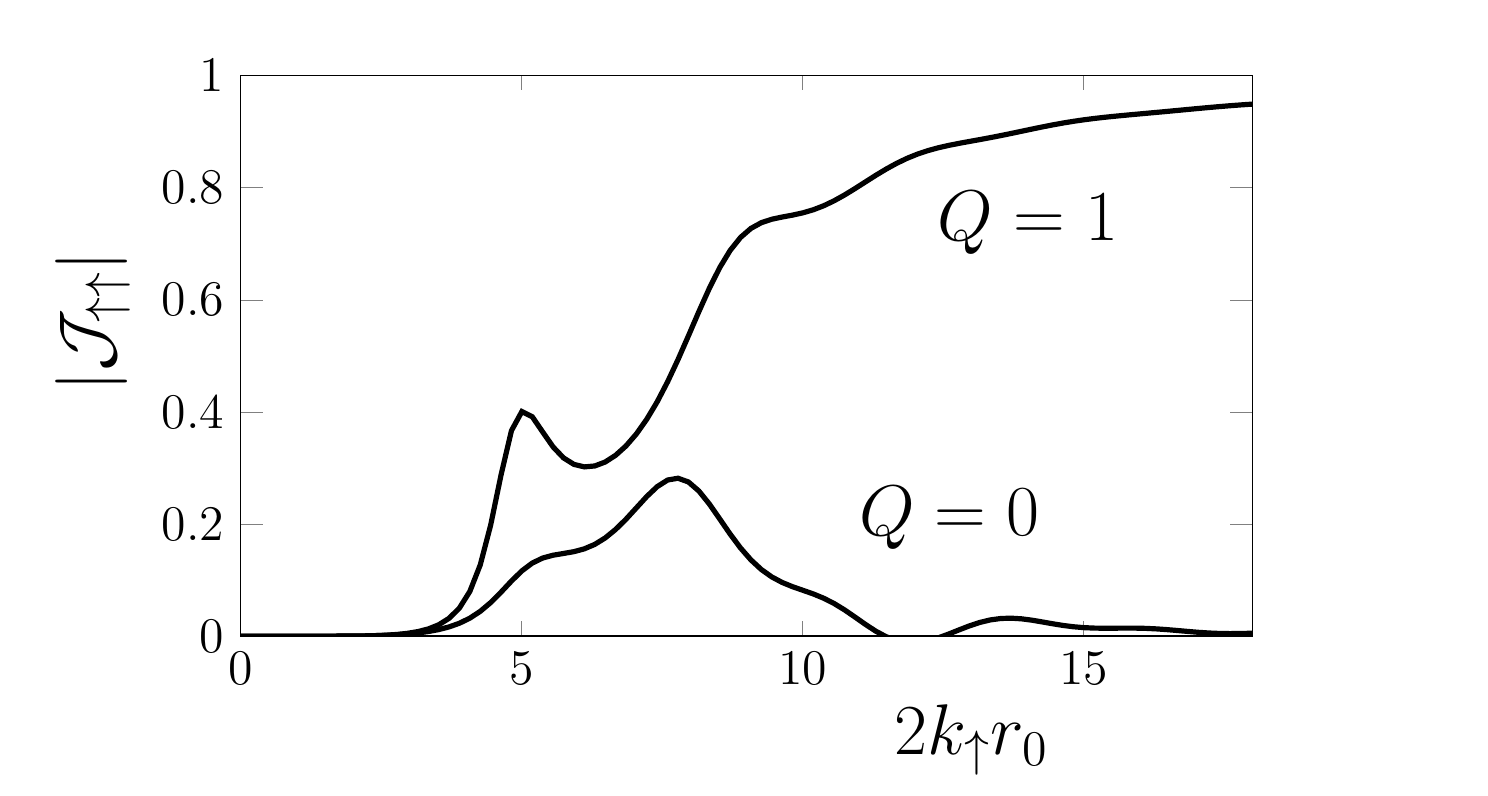}
	\caption{The dependence of the total asymmetric rate $\mathcal{J}_{\uparrow \uparrow}$ on the potential radius in units $kr_0$ for $Q=0,1$ spin configurations,
		the ratio $g/E = 2.5$.
	}	
	\label{f-one}
\end{figure}

We further apply this technique to study the scattering on a purely magnetic potential with the 
parametrization present in Eq.~\ref{eq_Skyrmion_Not}. 
In Fig.~\ref{f-one} we demonstrate the dependence of the total asymmetric rate $\mathcal{J}_{\uparrow \uparrow}$ 
on the magnetic texture radius $r_0$
for two topologically different configurations 
(in Fig.~\ref{f-one} the absolute value $|\mathcal{J}_{\uparrow \uparrow}|$ is shown; 
$\mathcal{J}_{\uparrow \uparrow}$ remains negative 
in accordance with the previous considerations). 
Here we take $\chi = 1$ and use the 
spin texture profiles from Eq.~\ref{eq_Q01}, the subband spin splitting $\Delta = 2g$. 
As follows from our calculations $\mathcal{J}_{\uparrow \uparrow}$ gets strongly suppressed at small $k r_0 \lesssim 1$. This fact has been already mentioned in~\ref{sWeak} when considering the perturbative region: 
it is due to the 
vanishing of the third-order correlator in Eq.~\ref{eq_Dissap}. 
We also notice that the asymptotic behavior of $\mathcal{J}_{\uparrow \uparrow}$ at $k_{\uparrow} r_0 \gg 1$ is similar to that observed for two opened spin subbands (see~\ref{s-Quasi} and Fig.~\ref{f3-7}). 
Namely, the asymmetric rate $\mathcal{J}_{\uparrow \uparrow} \to -1$ saturates for the topologicaly charged configuration $Q=1$ while going to zero for the uncharged one $Q=0$. 
This result stems from the Berry phase description~\ref{sAdiab} valid for each subband independently. 
It is worth mentioning, however, that 
in the intermediate region when 
neither perturbative theory nor the Berry phase approach are valid 
the asymmetric scattering 
generally persists 
independently of a spin texture topology. As follows from Fig.~\ref{f-one} in the range of $(5 \lesssim 2 k_{\uparrow} r_0 \lesssim 10)$ the scattering rate $\mathcal{J}_{\uparrow \uparrow}$ has approximately the same magnitude for $Q=0,1$ configurations. 


\section{Conclusions}
To summarize, we have considered the asymmetric electron scattering on a skyrmion-like magnetic texture. 
We have obtained a number of analytical results valid 
in the limiting regimes of weak and strong coupling, we have also developed a numerical scheme allowing to quantify the Hall response for an arbitrary case. 
The present analysis 
has revealed that 
when the electron orbital and spin motions 
can be viewed classically 
the magnitude of the Hall current is determined by the topological charge of a magnetic texture. 
However, we argue that beyond these conditions the topology of a magnetization is not relevant for the appearance of the Hall response.  
In particular, we have shown that 
in the weak coupling regime the asymmetric scattering rates have the same angular structure independently of a scattering potential profile. 
We have also analyzed the behavior of the scattering pattern in case of an electrically charged skyrmion. 
The presence of a short-range impurity is mostly important in the nonadiabatic regime, at that the charge transverse response due to magnetic texture is superseded by the spin Hall effect. 

\section*{Acknowledgments}
The Author thanks I.V.~Rozhansky and N.S.~Averkiev for  helpful and fruitful discussions. The work has been carried out with the financial support of the Russian Science Foundation (project 18-72-10111).
K.S.D. also thanks the Foundation for the Advancement of Theoretical Physics and Mathematics “BASIS”. 

\vspace{-.6cm}
\bibliography{Ref}

\end{document}